\documentclass[aip,superscriptaddress,twocolumn,eqsecnum,byrevtex]{revtex4}
\usepackage{amsfonts}
\usepackage{amssymb}
\usepackage{amsmath}
\usepackage{color}
\usepackage{graphicx}
\usepackage{time}
\usepackage{sidecap}
\usepackage{chngcntr}

\begin{document}
%
%
\title{Real-Space, Real-Time Approach to Quantum-Electrodynamical Time-Dependent Density Functional Theory}

\author{Justin Malave}
\affiliation{Department of Physics and Astronomy, Vanderbilt University, Nashville, Tennessee, 37235, USA}
\author{Alexander Ahrens}
\affiliation{Department of Physics and Astronomy, Vanderbilt University, Nashville, Tennessee, 37235, USA}
\author{Daniel Pitagora}
\affiliation{Department of Physics and Astronomy, Vanderbilt University, Nashville, Tennessee, 37235, USA}
\author{Cody Covington}
\affiliation{Department of Chemistry, Austin Peay State University, Clarksville, USA}

\author{K\'alm\'an Varga}
\email{kalman.varga@vanderbilt.edu}
\affiliation{Department of Physics and Astronomy, Vanderbilt University, Nashville, Tennessee, 37235, USA}

\begin{abstract}
The Quantum-Electrodynamical Time-Dependent Density Functional Theory
(QED-TDDFT) equations are solved by time propagating the wave function
on a tensor product of a Fock-space and real-space grid. Applications
for molecules in cavities show the accuracy of the approach. Examples
include the coupling strength and light frequency dependence of the
energies, wave functions, optical absorption spectra, and Rabi
splitting magnitudes in cavities, as well as a description of high harmonic
generation in cavities.

\end{abstract}

\maketitle
\section{Introduction}
The possibility of altering physical and chemical properties 
by
coupling matter to light has attracted 
intense experimental 
\cite{https://doi.org/10.1002/anie.201107033,
Balili1007,PhysRevLett.114.196403,Xiang665,
doi:10.1021/acsphotonics.0c01224,Coles2014,Kasprzak2006,PhysRevLett.106.196405,Plumhof2014,
https://doi.org/10.1002/adma.201203682,Wang2021,BasovAsenjoGarciaSchuckZhuRubio+2021+549+577} 
and theoretical interest\cite{PhysRevLett.114.196402,PhysRevLett.121.253001,
PhysRevLett.128.156402,Riso2022,
https://doi.org/10.1002/qua.26750,
PhysRevLett.122.017401,DiStefano2019,PhysRevX.5.041022,
PhysRevLett.116.238301,Galego2016,Shalabney2015,doi:10.1021/acsphotonics.9b00648,Schafer4883,
Ruggenthaler2018,Flick15285,Flick3026,PhysRevLett.123.083201,Mandal,doi:10.1021/acsphotonics.9b00768,
doi:10.1021/acs.nanolett.9b00183,FlickRiveraNarang,PhysRevLett.121.113002,Garcia-Vidaleabd0336,
Thomas615,PhysRevResearch.2.023262,doi:10.1063/5.0036283,doi:10.1063/5.0038748,doi:10.1063/5.0039256,
doi:10.1063/5.0012723,doi:10.1063/5.0021033,acs.jpcb.0c03227,PhysRevLett.119.136001,
doi:10.1063/5.0012723,Flick15285,doi:10.1021/acsphotonics.7b01279,PhysRevA.98.043801,
doi:10.1021/acs.jpclett.0c01556,doi:10.1021/acs.jctc.0c00618,doi:10.1021/acs.jpclett.0c03436,
doi:10.1021/acs.jctc.0c00469,PhysRevB.98.235123,PhysRevLett.122.193603,
Szidarovszky_2020,doi:10.1021/acs.jpclett.1c01570,doi:10.1021/jacs.2c00921,
doi:10.1063/5.0095552,doi:10.1021/acs.jpclett.1c02659,doi:10.1021/jacs.1c13201,
Cederbaum2021}. 
{There are several excellent review articles highlighting the
present state of the art of experimental and theoretical approaches related
to light-matter interaction in cavities. These include reviews about
the  properties of hybrid light-matter states
\cite{doi:10.1021/acs.accounts.6b00295,doi:10.1063/PT.3.4749}, ab
initio calculations \cite{Ruggenthaler2018,doi:10.1063/5.0094956} and
molecular polaritonics
\cite{doi:10.1146/annurev-physchem-090519-042621,doi:10.1021/acsphotonics.2c00048,
doi:10.1021/acsphotonics.1c01749}.}

The theoretical
and computational description of the coupled light matter system is
challenging because the already difficult quantum many-body problem of 
electrons and nuclei is aggravated with the addition of the photon degrees of freedom. 
A plethora of approaches going beyond the simple two-level atom model
\cite{Jaynes1962ComparisonOQ} has been proposed in the last few years. Most of these
approaches are based on successful many-body quantum methods adapted to include photon interactions
. The use of the Pauli-Fierz (PF) non-relativistic quantum electrodynamics Hamiltonian  
is found to be the most useful framework 
\cite{Ruggenthaler2018,Rokaj_2018,Mandal,acs.jpcb.0c03227,PhysRevB.98.235123}
for practical calculations. The PF Hamiltonian is a sum of electronic and
photonic Hamiltonians and a cross term describing the electron-photon
interaction. {Due to this cross term, one has to use a coupled
electron-photon wave function, 
\begin{equation}
\sum_{\vec{n}} \Phi_{\vec{n}}(\mathbf{x})\chi_{\vec{n}}
\label{coupled}
\end{equation}
where $\mathbf{x}=(\mathbf{r}_1,\mathbf{r}_2,{\ldots},
\mathbf{R}_1,\mathbf{R}_2,{\ldots})$ are the spatial coordinates of the electrons and
nuclei, and $\vec{n}=(n_1,n_2,{\ldots} N_p)$ are the quantum numbers of the photon
modes. The occupation number basis, $\chi_{\vec{n}}=\vert
n_1,n_2,{\ldots},N_p\rangle$, 
is used to represent the bosonic Fock-space of photon
modes (see Appendix \ref{boson} for a more detailed definition). }

Wave function based approaches
\cite{PhysRevLett.127.273601,doi:10.1063/5.0066427,doi:10.1063/5.0078795,doi:10.1063/5.0038748,PhysRevX.10.041043}
typically use coupled electron-photon wave functions and the product
form significantly increases the dimensionality. In Refs.
\cite{doi:10.1063/5.0078795,doi:10.1063/5.0038748,PhysRevX.10.041043,
PhysRevResearch.2.023262} the 
coupled cluster (CC) approach is used by defining a reference wave
function as a direct product of a
Slater determinant of Hartree-Fock states and a zero-photon photon
number state. The exponentiated cluster operator acting upon this
product state defines the ground state QED-CC wave function. 
Systematic improvability is the greatest benefit of this approach. 
The stochastic variational method
\cite{PhysRevLett.127.273601,doi:10.1063/5.0066427} (QED-SVM)
also uses a product of matter and photonic wave functions but in this
case, the matter part is described by explicitly correlated Gaussian
basis states 
\cite{RevModPhys.85.693}. The parameters of the variational ansatz 
are optimized by a stochastic selection process leading to highly
accurate energies and wave functions. Both the QED-CC and QED-SVM 
approaches are limited  to small atoms and molecules.

{A pioneering approach to describe the
interaction of light and matter in cavities  is the quantum electrodynamical density functional
theory (QEDFT) \cite{doi:10.1063/5.0039256,doi:10.1021/acsphotonics.7b01279,
PhysRevA.98.043801,PhysRevA.90.012508,PhysRevLett.110.233001,doi:10.1063/5.0021033,doi:10.1063/5.0057542}.
The  QEDFT is an exact reformulation of the PF Hamiltonian based 
many-body wave theory. In practical applications of QEDFT one has to develop
good approximations of the fields and currents so that the
auxiliary non-interacting system generates the same physical quantities 
as the interacting system. To facilitate this need, the development of polaritonic
exchange-correlation functionals is a focus of intense research interest
\cite{doi:10.1021/acsphotonics.7b01279,PhysRevLett.115.093001,
doi:10.1073/pnas.2110464118,https://doi.org/10.48550/arxiv.2104.06980}.
In QEDFT the spatial and photon wave functions are factorized 
allowing the separation of the electronic and photonic
components. One then has to solve coupled equations for the
matter and photon parts.  The dimensionality limits the
product basis approaches to a few photon modes, while the QEDFT can be
applied to hundreds of thousands of photon modes
\cite{doi:10.1021/acsphotonics.9b00768}. More details and applications
of the QEDFT method and its time-dependent version can be found 
in Refs. \cite{Schafer4883,Ruggenthaler2018,Flick3026,
doi:10.1021/acs.jpclett.0c03436,doi:10.1021/acsphotonics.9b00768}.}

In this paper, we present a QED-DFT (and QED-TDDFT) approach using a coupled
electron-photon wave function like that shown in  Eq. \eqref{coupled}. 
The wave function will be defined on a tensor product (TP) of a spatial grid and
Fock state representation.  To distinguish this approach from the aforementioned QEDFT approach, 
we will call the present approach QED-DFT-TP and QED-TDDFT-TP. {The QED-DFT-TP is a particular realization
of the QEDFT theory using a different ansatz -- a coupled
electron-photon wave function. }

The tensor product form increases the dimensionality; however, it preserves the quantized photon states. In this
way, we  have direct access to non-classical observables of the photon field
e.g., the photon-number, the purity, or the Mandel Q number \cite{Mandel:79}.
The coupled electron-photon wave function provides a more complex
description of the light-matter interaction by calculating the spatial
wave function in each photon sector. 

Each orbital of the molecule is coupled
with different Fock basis states (photon states) describing the quantized light. The
light-matter coupling part of the Hamiltonian will describe the interaction between
the orbital components in different photons states. The Fock states
are orthogonal, preserving and even  increasing the sparsity of the
real-space based DFT Hamiltonian. This sparsity allows the use of the
efficient iterative diagonalization approaches 
traditionally used in real-space DFT approaches.

As the coupling part and the product ansatz are simple, the
present approach can be easily implemented in any real-space
approaches. Plane-wave or orbital-based DFT methods can see similarly
simple adaptation.

The approach will be tested  using time-dependent and time-independent problems
and our results will be compared to QED-SVM, QED-CC, and QEDFT calculations. One expects that
the approach can be used everywhere where regular DFT and TDDFT have
been useful, from bond length and density distribution calculations, to
descriptions of high harmonic generation and optical absorption.

\section{Formalism}
\subsection{Hamiltonian}
We assume that the system is nonrelativistic and the coupling 
to the light can be described by the dipole approximation.  
The Pauli-Fierz non-relativistic QED Hamiltonian 
provides a consistent quantum description at this level 
\cite{Ruggenthaler2018,Rokaj_2018,Mandal,acs.jpcb.0c03227,PhysRevB.98.235123}. 
The  dipole approximation assumes that the spatial variation  of
the electric field is negligible across the size of the system -- 
physically valid if the system size is much smaller than the wavelength
of the light. The PF Hamiltonian in the Coulomb gauge is
$H=H_e+H_{ep}$
where $H_e$ is the electronic Hamiltonian and
\begin{eqnarray}
H_{ep}&=&
\sum_{\alpha=1}^{N_p}{1\over 2}\left[p_{\alpha}^2+ 
\omega_{\alpha}^2
\left(q_{\alpha}-{\boldsymbol{\lambda}_{\alpha}\over\omega_{\alpha}}\mathbf{D}\right)^2\right]
\\
&=&
\sum_{\alpha=1}^{N_p}\left[ \omega_{\alpha}\left(\hat{a}_{\alpha}^{+} \hat{a}_{\alpha}+\frac{1}{2}\right)
-\omega_{\alpha}q_\alpha
    \boldsymbol{\lambda}_{\alpha}\mathbf{D}+{\frac{1}{2}}  \left(\mathbf{\lambda_{\alpha}}
\mathbf{D}\right)^{2}\right], \nonumber
\label{hep}
\end{eqnarray}
(atomic units are used in this work). In Eq. \eqref{hep}  $\mathbf{D}$ is the dipole operator, 
the photon fields are described by quantized oscillators, and
$q_\alpha={\frac{1}{ \sqrt{2\omega_\alpha}}}(\hat{a}^+_\alpha+\hat{a}_\alpha)$ 
is the displacement field. Here $\hat{a}$ and $\hat{a}^+$ are,
respectively, the
lowering and raising operators of the quantized harmonic oscillator.
This  Hamiltonian describes $N_p$ photon modes with photon frequency
$\omega_{\alpha}$ and coupling  $\boldsymbol{\lambda}_{\alpha}$.  
The coupling term is written as \cite{PhysRevA.90.012508} 
$\boldsymbol{\lambda}_{\alpha}={1/\sqrt{\epsilon_0}}\mathbf{u}_\alpha(\mathbf{r}_0)$,
where $\mathbf{u}_\alpha(\mathbf{r})$ is the cavity mode function and
$\mathbf{r}_0$ is the center of the cavity where the molecule is placed.
The first term in Eq. \eqref{hep} is the Hamiltonian of the photon modes, the second 
term couples the photons to the dipole, and the last term is the dipole self-interaction,
$$
H_d={1\over 2} \sum_{\alpha=1}^{N_p}\left(\boldsymbol{\lambda}_{\alpha} \mathbf{D}\right)^{2}.
$$

For the electronic Hamiltonian we adapt the Kohn-Sham (KS) TDDFT 
\cite{PhysRevLett.52.997} description,
\begin{equation}
H_e=-{\hbar^2\over 2m }\nabla^2  + V_{KS}(\mathbf{r}),
\label{tdks}
\end{equation}
with
\begin{equation}
V_{KS}(\mathbf{r})=V_{\mathrm{H}}[\rho](\mathbf{r})+V_{\mathrm{XC}}[\rho](\mathbf{r})
+V_{\mathrm{ion}}(\mathbf{r}).
\end{equation}
Here, $\rho$ is the electron density, $V_\mathrm{H}$ is the Hartree potential, 
and  $V_\mathrm{XC}$ is the exchange-correlation potential (Local
density approximation (LDA) is used), and
$V_{\mathrm{ion}}$ is the external potential due to the ions. The potential of the ions can be
represented by employing the norm-conserving pseudopotentials of the form given by Troullier and Martins
\cite{troullier_prb_43_1993_1991}. We assume that the nuclei are fixed
in their positions.  

{One should emphasize at this point that the proper way of
introducing the Hamiltonian would be to use mapping theorems following the steps 
of Ref. \cite{Flick15285,PhysRevLett.115.093001,PhysRevA.90.012508}. See also 
Ref. \cite{doi:10.1073/pnas.2110464118} for recent developments. Our
approach can be thought to be using the conventional velocity gauge TDDFT Hamiltonian
as a starting point
\begin{equation}
H_V={1\over 2 }\left(-i\hbar \nabla-{e\over c} \mathbf{A}\right)^2  + V_{KS}(\mathbf{r}).
\end{equation}
Then we replace the classical vector potential with the
quantized vector potential of the cavity:
\begin{equation}
\hat{\mathbf{A}}(\mathbf{r})=
\sum_{\alpha} \left({\hbar\over 2\epsilon_0 \omega_\alpha}\right)^{1/2}
\left[
\hat{a}_{\alpha}\mathbf{u}_{\alpha}(\mathbf{r})+
\hat{a}_{\alpha}^+\mathbf{u}_{\alpha}^*(\mathbf{r})
\right].
\end{equation}
In the dipole approximation, placing the molecule in the middle of the
cavity at $\mathbf{r}_0$ one has
\begin{equation}
\hat{\mathbf{A}}(\mathbf{r}_0)=
\sum_{\alpha} \boldsymbol{\lambda}_{\alpha}\hat{a}_{\alpha},
\end{equation}
and
\begin{equation}
H_V={1\over 2 }\left(-i\hbar \nabla-{e\over c}
\hat{\mathbf{A}}(\mathbf{r}_0)\right)^2  + V_{KS}(\mathbf{r}).
\end{equation}
One can use a unitary transformation to transform the Hamiltonian into
length gauge \cite{Rokaj_2018,Taylor:22}
\begin{equation}
H_L=U^{\dagger}H_VU,
\end{equation}
with 
\begin{equation}
U=\exp\left\lbrace-{i\over \hbar} 
\hat{\mathbf{A}}(\mathbf{r}_0)\cdot \mathbf{D}\right\rbrace.
\end{equation}
After a straightforward calculation \cite{Rokaj_2018} one has
\begin{equation}
H_L=-{\hbar^2\over 2m }\nabla^2  + V_{KS}(\mathbf{r})+
\sum_{\alpha=1}^{N_p}{1\over 2}\left[p_{\alpha}^2+ 
\omega_{\alpha}^2
\left(q_{\alpha}-{\boldsymbol{\lambda}_{\alpha}\over\omega_{\alpha}}\mathbf{D}\right)^2\right].
\label{tdks1}
\end{equation}
Either $H_V$ or $H_L$ can be used in the calculations. The advantage
of $H_V$ is that one can use it with periodic boundary conditions. In
either case, the Hamiltonian depends not only on the spatial
coordinates, but on the photon creation and annihilation operators, as
well and one has to use a tensor product of a spatial and number state
basis for the orbitals. We will use $H=H_L$ in this work.
}

\subsection{Basis functions}
The $\omega_{\alpha}q_\alpha\boldsymbol{\lambda}_{\alpha}\mathbf{D}$ term
couples the electronic and photonic degrees of freedom. The orbitals
of the coupled electron and photon system at the KS level  can be written 
\begin{equation}
\Phi_{mn}=\phi_{mn}(\mathbf{r})|n\rangle,  
\ \ \ \ \  (m=1,{\ldots},N_{occ}),
\ \ \ \ \  (n=0,{\ldots},N_F),
\end{equation}
where $|n\rangle$ is the Fock space basis for the photons, $N_F$ is
the dimension of the Fock space, and $N_{occ}$ is the number of
orbitals. In the
following we assume that there is one dominant photon mode -- $N_p=1$ in
Eq. \ref{hep} -- with frequency $\omega$. Extension to $N_p>1$ is
possible, but it quickly leads to prohibitively large basis dimensions.  
We use a real-space  representation 
\cite{PhysRevLett.72.1240,PhysRevB.54.4484,doi:10.1063/1.5142502,PhysRevLett.93.176403}
for the electronic part 
\begin{equation}
\phi_{mn}(\mathbf{r})=\phi_{mn}(x,y,z),
\end{equation}
and the orbitals will be defined on a 4 dimensional (4D) product grid of
the 3 dimensional real-space grid and the 1 dimensional Fock space.
The 4D grid has $N_x\times N_y \times N_z\times (N_F+1)$ grid points, where
$N_x,N_y,N_z$ are the grid points in real-space and $N_F$ is the size
of the Fock basis. The orbitals are written as
\begin{equation}
\Phi_{m}(\mathbf{r})=\left(
\begin{array}{c}
\Phi_{m0}(x_i,y_j,z_k)\\
\Phi_{m1}(x_i,y_j,z_k)\\
\vdots\\
\Phi_{mn}(x_i,y_j,z_k)\\
\vdots
\end{array}
\right).
\end{equation}
Due to the orthogonality of the Fock basis states we have
\begin{equation}
\left(\Phi_{mn}\vert\Phi_{m'n'}\right)=
\langle\phi_{mn}\vert\phi_{m'n}\rangle \delta_{nn'},
\end{equation}
where the round bracket stands for the integration over the real and
Fock space and the angle bracket is the integration over the space part,
\begin{equation}
\langle\phi_{mn}\vert\phi_{m'n}\rangle =\sum_{ijk}
\phi_{mn}(x_i,y_j,z_k) \phi_{m'n}(x_i,y_j,z_k).
\end{equation}
Using Gram-Schmidt orthogonalization the real-space functions
$\phi_{1n},{\ldots},\phi_{N_{occ}n}$ are orthogonalized for each Fock
state $(n=0,{\ldots},N_F)$. The new orthogonal set
$\hat{\phi}_{1n},{\ldots},\hat{\phi}_{N_{occ}n}$ is normalized
\begin{equation}
\sum_{n=0}^{N_F}\vert\hat{\phi}_{mn}\vert^2=1.
\end{equation}
Now one can define the electron density as
\begin{equation}
\rho(\mathbf{r})=\sum_{m=1}^{N_{occ}} c_m
\sum_{n=0}^{N_F}\vert\hat{\phi}_{mn}(\mathbf{r})\vert^2,
\end{equation}
where $c_m$ is the number of electrons on the $m$th orbital. One can
also define the density belonging to a given Fock state as
\begin{equation}
p_n(\mathbf{r})=\sum_{m=1}^{N_{occ}} c_m \vert\hat{\phi}_{mn}(\mathbf{r})\vert^2,
\end{equation}
and the photon occupation probability in the Fock space 
\begin{equation}
P_n={1\over N}\int p_n(\mathbf{r}) d{\mathbf r},
\end{equation}
where $N$ is the number of electrons.

The orbitals will be calculated by iterative minimization -- conjugate
gradient in the present case -- in
the same way as in the conventional real-space approaches. For this we
need to calculate the action of the Hamiltonian on the wave function.
Noting that 
\begin{equation}
\omega\left(\hat{a}^+\hat{a}+{1\over 2}\right)
\vert n\rangle=(n+{1\over 2} )\omega\vert n\rangle,
\end{equation}
and
\begin{eqnarray}
q|n\rangle&=&\frac{1}{\sqrt{2
\omega}}\left(\hat{a}+\hat{a}^{+}\right)|n\rangle\\
&=&
\frac{1}{\sqrt{2 \omega}}\left(|\sqrt{n}|
n-1\rangle+\sqrt{n+1}|n+1\rangle\right) \nonumber,
\end{eqnarray}
one has
\begin{widetext}
\begin{equation}
H \Phi_{mn}(\mathbf{r})=-{1\over 2}\nabla^2
\Phi_{mn}(\mathbf{r})+\left(V_{KS}(\mathbf{r})+\mu(\boldsymbol{\lambda}\mathbf{r})+
\omega\left(n+{1\over2}\right)\right)\Phi_{mn}(\mathbf{r})
-\sqrt{\omega\over
2}(\boldsymbol{\lambda}\mathbf{r})\left(\sqrt{n}\Phi_{mn-1}(\mathbf{r})+
\sqrt{n+1}\Phi_{mn+1}(\mathbf{r})\right),
\label{hact}
\end{equation}
\end{widetext}
where 
\begin{equation}
\mu=\int \boldsymbol{\lambda}\mathbf{r} \rho(\mathbf{r})d\mathbf{r}.
\end{equation}
In the first part of this equation, the Hamiltonian acts on
$\phi_{mn}$ in each Fock space in the same way as in conventional 
real-space approaches.
The kinetic energy operator is represented by nine-point finite
differencing, and the nonlocal part of the pseudopotential is calculated
by a summation in the pseudopotential core radius around the atomic position, leading to
a very sparse Hamiltonian matrix.  In the second part of the equation 
the photon spaces $n$ and $n\pm 1$ are connected and the coupling
Hamiltonian matrix is diagonal. This large sparse system is ideal for
iterative approaches. 

\subsection{Ground state calculation}
The ground state calculation is similar to the conventional DFT 
calculations:
\begin{enumerate}
\item
Initialization of the orbitals. $\phi_{mn}^{(0)}$ are approximated 
e.g. by atomic
orbitals. The components belonging to the lowest ($n=0,1$) Fock spaces
are expected to be dominant so a weight factor is used to enhance those
components, $\phi^{(0)}_{mn} \rightarrow w_n\phi^{(0)}_{mn}$.
Orthogonalize the orbitals to generate the starting wave function 
\begin{equation}
\hat{\Phi}_{m}^{(0)}(\mathbf{r})=\left(
\begin{array}{c}
\hat{\Phi}^{(0)}_{m0}(\mathbf{r})\\
\hat{\Phi}^{(0)}_{m1}(\mathbf{r})\\
\vdots\\
\hat{\Phi}^{(0)}_{mn}(\mathbf{r})\\
\vdots
\end{array}
\right),
\end{equation}
where $\hat{\Phi}^{(0)}_{mn}(\mathbf{r})=\hat{\phi}^{(0)}_{mn}(\mathbf{r})\vert n
\rangle$.
\item
Iterative minimization step. Use
\begin{equation}
{\Phi}_{m}^{(k+1)}(\mathbf{r})=\sum_j a_j H^j
\hat{\Phi}_{m}^{(k)}(\mathbf{r}),
\end{equation}
where the $a_j$ coefficients define the iterative procedure, e.g.
steepest descent, conjugate gradient, imaginary time-propagation, etc.
Eq. \ref{hact} is used to calculate $H\hat{\Phi}^{(k)}_{mn}$.
\item 
Calculate $\hat{\Phi}_{m}^{(k+1)}(\mathbf{r})$ by Gram-Schmidt
orthogonalization 
${\Phi}_{m}^{(k+1)}(\mathbf{r})$.
\item 
Calculate $\rho$ and update $V_{KS}$. 
\item
k=k+1, go to step 2 until convergence is reached.
\end{enumerate}

\subsection{Time propagation}
The ground state orbitals will be used to initialize the time
propagation
\begin{equation}
\hat{\Phi}_{m}(\mathbf{r},t=0)=
\hat{\Phi}_{m}(\mathbf{r}).
\end{equation}
Any time propagation method typically used for TDDFT can be used here, and in this work
we use Taylor time propagation \cite{PhysRevB.54.4484}
\begin{equation}
\hat{\Phi}_{m}(\mathbf{r},t+\Delta t)={\rm e}^{-iH\Delta t}
\hat{\Phi}_{m}(\mathbf{r},t)=\sum_{j=0}^4 {(-i\Delta t)^j\over j!}
H^j\hat{\Phi}_{m}(\mathbf{r},t),
\end{equation}
where $\Delta t$ is chosen to be sufficiently small to conserve the norm of the orbitals
during propagation. 

\section{Applications}
Atomic units are used in the calculations, and we drop the ``a.u.'' notation
after every number. The coupling will be
defined as $\boldsymbol{\lambda}=\lambda \mathbf{u}$ where $\mathbf{u}$
is a unit vector, e.g. (1,0,0). $\lambda$ can be related to the
effective cavity volume $\lambda=1/\sqrt{\epsilon_0 V_{eff}}$
($\epsilon_0$ is the permittivity of vacuum) \cite{PhysRevX.9.021057}.
Sub nm$^3$ volumes have been reached in picocavities \cite{pico1,pico2} 
corresponding to $\lambda<0.05$, which will be the highest value that
we consider in this work, except for an example for benzene in which 
we calculate the wave function in higher photon spaces.

\subsection{Hydrogen molecule}
\begin{figure}
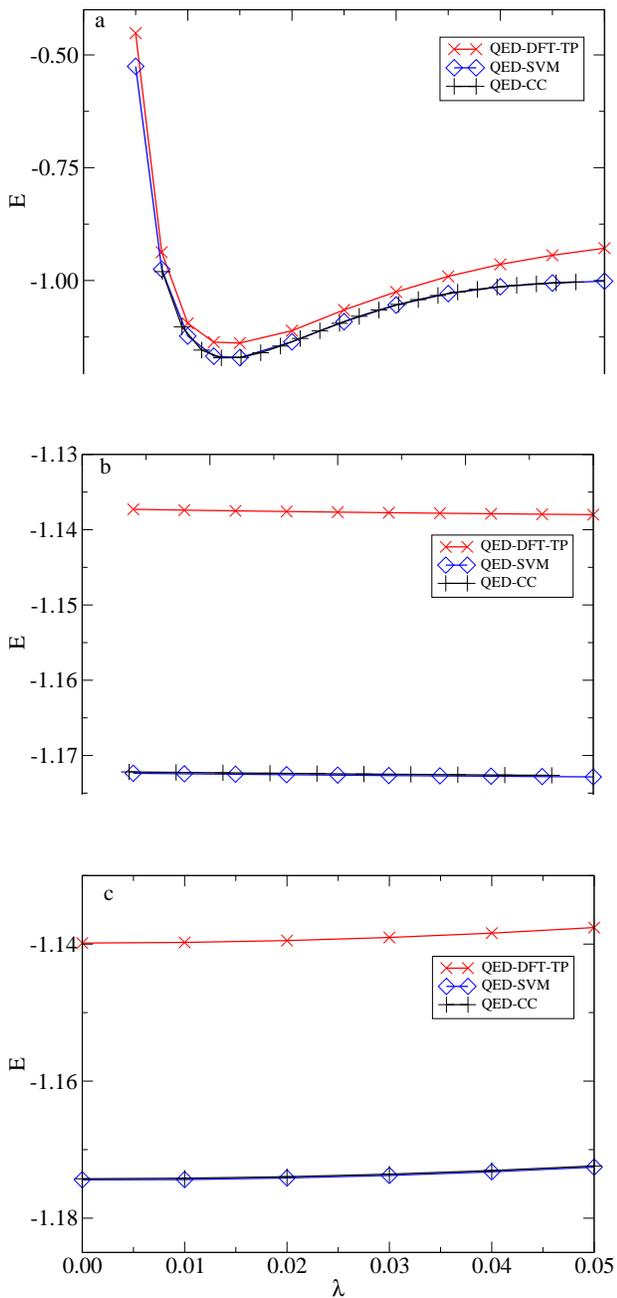

\includegraphics[width=0.45\textwidth]{figure1a.eps}
\\
\includegraphics[width=0.45\textwidth]{figure1b.eps}
\\
\includegraphics[width=0.45\textwidth]{figure1c.eps}
\caption{Comparison of the results of the QED-DFT-TP, QED-CC and QED-SVM
calculations for energy dependence on bond length (a), $\omega$ (b)
and $\lambda$(c) . 
The QED-CC results are from the calculation of Ref. \cite{rd} using
cc-pVQZ basis. $\omega=0.08$ is used for a and c,
$\lambda$=0.05 is used for a and b. $N_x=N_y=N_z=81$ and $h=0.5$ grid
spacing used in the calculations. }		
\label{h2}
7\end{figure}
In the first example, we calculate the energy of a hydrogen molecule in
a cavity with fixed nuclei. Two-photon spaces, $|0\rangle$ and $|1\rangle$,
are used. Higher photon spaces have negligible contribution
for this coupling strength region.  The results are compared to an SVM and a
QED-CC calculation. As mentioned in the introduction, both the SVM
and the QED-CC are very accurate wave function based approaches and
are in complete agreement with each other in these calculations.
Fig. \ref{h2} shows the total energy as a function of bond length
(denoted as $d$) for
$\lambda=0.05$ ($\boldsymbol{\lambda}=(\lambda,0,0))$ and $\omega=0.081$. 
The molecular bond lay parallel to the photon polarization vector. The
parameters are chosen to make a comparison to the QED-CC calculations
of Ref. \cite{rd}.
The calculated QED-DFT-TP energy is shifted with respect to QED-CC and SVM energies, 
due to the LDA and pseudopotential approximations, but the overall
trend is very similar. The only noticeable difference is in the 2 $\le$ $d$
region where the QED-DFT-TP curve has a larger slope. This is mainly due
to the LDA approximation which performs well near the equilibrium
separation, but inaccurate when the bond is stretched  as it is 
discussed e.g. in Ref. \cite{doi:10.1063/1.4869598}.

At this coupling strength, the effect of the
coupling on the ground state is mainly an energy shift. It is more
interesting to study the $\omega$ and $\lambda$ dependence for a given
bond length. This is shown in Figs \ref{h2} b and c. The SVM and
QED-CC results are in excellent agreement, and the QED-DFT-TP results
show the same behavior except for the energy shift. 

\begin{figure}
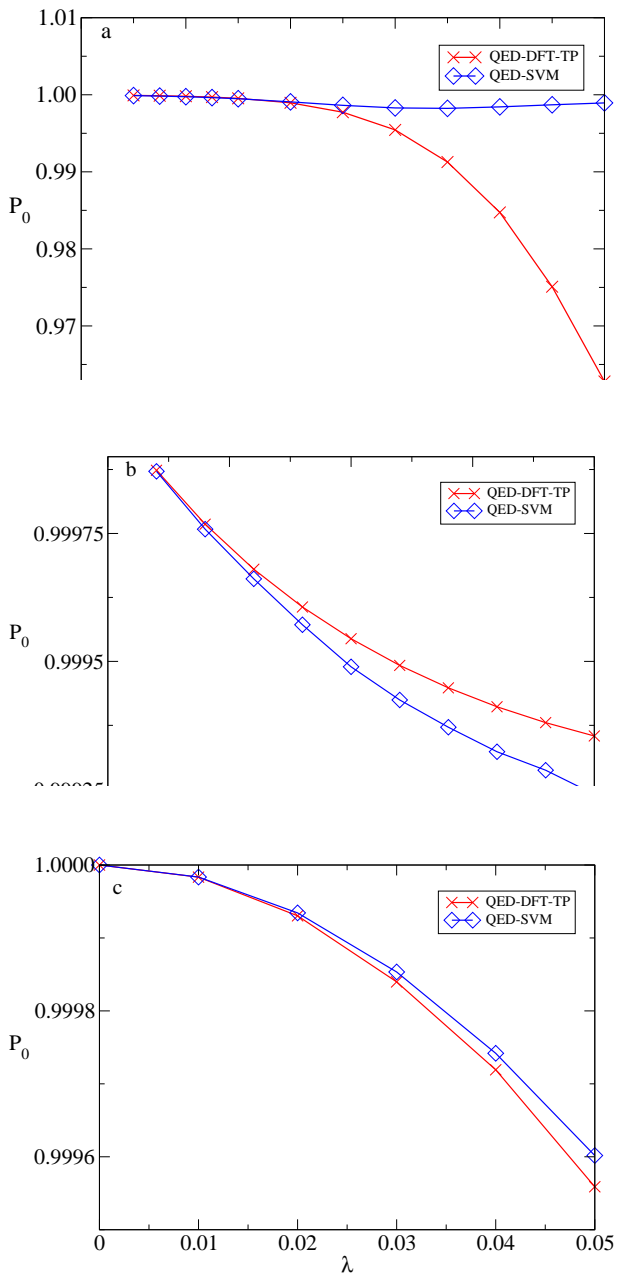

\includegraphics[width=0.45\textwidth]{figure2a.eps}\\
\includegraphics[width=0.45\textwidth]{figure2b.eps}\\
\includegraphics[width=0.45\textwidth]{figure2c.eps}
\caption{Photon occupation number as a function of bond length (a), $\omega$ (b)
and $\lambda$(c) . 
The QED-CC results are from \cite{rd}. $\omega=0.08$ is used for a and c,
$\lambda$=0.05 is used for a and b. $N_x=N_y=N_z=81$ and $h=0.5$ used
in the calculations. }		
\label{h2p}
\end{figure}

Fig. \ref{h2p} compares photon occupation number $P_n$ as a function of
bond length $\lambda$ and $\omega$. $N_F=1$ is used, so $P_0+P_1=1$
and only $P_0$ is shown in the figures. In the case of bond length, SVM and QED-DFT agree
well up to $d=1.7$. For larger distances, the two
probabilities are different, probably due to the aforementioned LDA issue. 
The SVM and QED-DFT-TP description of the $\lambda$ and $\omega$ 
dependence agrees well.

\subsection{The LiH and the HF molecules} 
\begin{figure*}
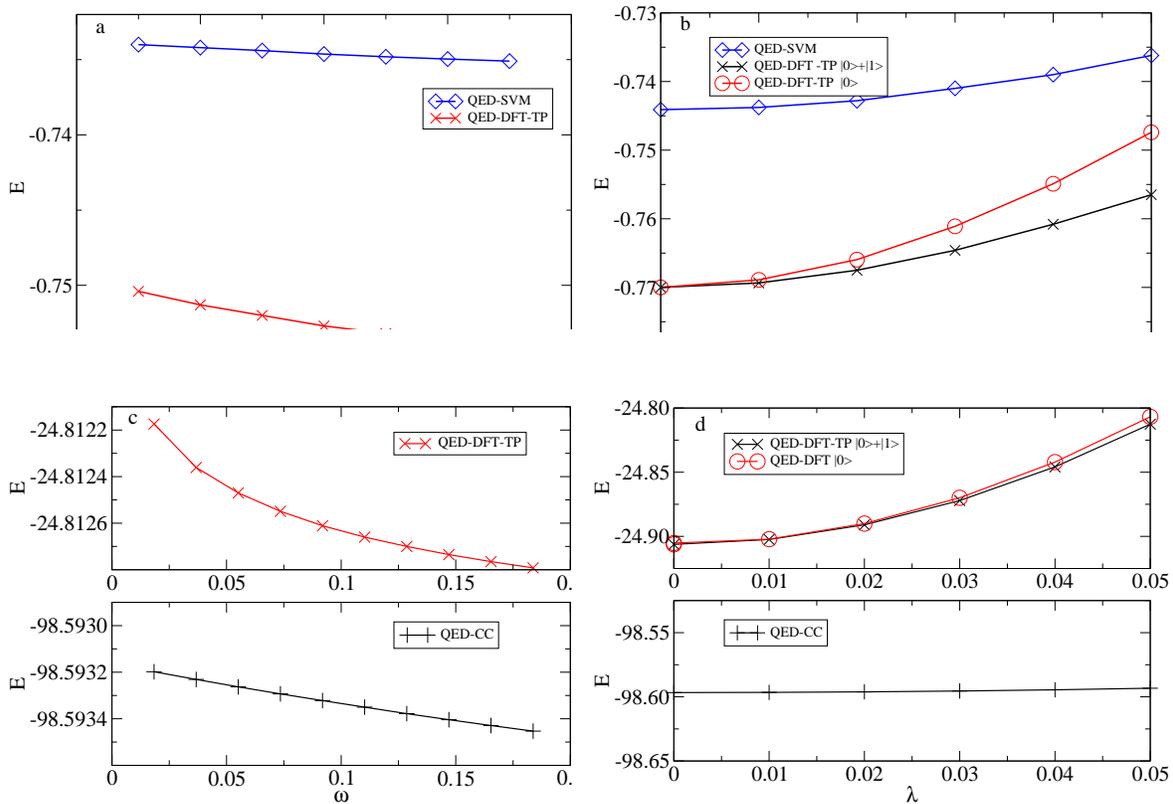

\begin{minipage}{0.95\textwidth}
\includegraphics[width=0.45\textwidth]{figure3a.eps}
\includegraphics[width=0.45\textwidth]{figure3b.eps}
\includegraphics[width=0.45\textwidth]{figure3c.eps}
\includegraphics[width=0.45\textwidth]{figure3d.eps}
\end{minipage}
\caption{Dependence of the energy of LiH and HF molecules on $\omega$
and $\lambda$. The Li-H distance is 3.015 and the 
H-F distance is 1.7229, $\lambda=0.05$ was used to calculate the
$\omega$ dependence for both molecules. For LiH $\omega=0.5$, for HF
$\omega=0.0813$ was used to calculate the $\lambda$ dependence. 
$N_x=N_y=N_z=51$ grid points is used with $h=0.5$ grid
spacing. The QED-CC results are from \cite{rd}.
In the case of the LiH molecule the QED-SVM results are shifted up by
7.3.}	      
\label{hf}
\end{figure*}

Figs. \ref{hf}a and b shows the $\lambda$ and $\omega$ dependence of
the energy of the LiH molecule. The $\omega$ dependence calculated
by QED-SVM and QED-DFT-TP are very similar. Due to the pseudopotential
approach the total energies calculated by QED-DFT-TP and QED-SVM are
different and the QED-SVM energies are shifted up by a constant
to fit them in the same figure. We chose a  larger $\omega$ ($\omega=0.5)$ for the
calculation of the $\lambda$ dependence to show the importance of the
coupling of the $|0\rangle$ and $|1\rangle$ photon spaces. Fig.
\ref{hf}b shows that the energy in the $|0\rangle$+$|1\rangle$ case is
significantly lower than in the $|0\rangle$ case for higher $\lambda$
values (in the $|0\rangle$ space only the dipole self interaction is added
to the Hamiltonian). The $\lambda$ dependence calculated by 
QED-SVM and the QED-DFT-TP are in good agreement.

Figs. \ref{hf}c and d shows the $\lambda$ and $\omega$ dependence for a HF
molecule. The $\omega$ dependence is very similar in both the
QED-DFT-TP and QED-CC cases. The energy shift is due to the pseudopotential
treatment of the core electrons in the QED-DFT-TP case. 
In this case the weight of the $|0\rangle$ space is 95\% and 
the weight of the $|1\rangle$ space is 5\% for the largest $\lambda$,
so in overall the $|0\rangle$ space dominates but the $|1\rangle$ space 
also contributes to the energy as Fig. \ref{hf}d shows. 

The $\lambda$ dependence only agrees in the tendency that by increasing $\lambda$ the
energy is increased, but the $\lambda$ dependence is much larger in
the QED-DFT-TP case.

\subsection{Na$_{2}$ cluster}
The next example is a Na$_2$ cluster with a bond length $d$=5.8. After the ground
state calculation the initial wave function is excited with a delta
kick (see Appendix \ref{abs} for details) and then time propagated. The
photon coordinate $q(t)$ and the dipole moment $D_x(t)$ are shown in
Fig. \ref{Na2}. We have calculated these quantities by the QED-TDDFT-TP
and with the QEDFT approaches (see
Appendix \ref{Rubio} for details). The time dependence of $q(t)$
calculated by the two methods differ greatly. This is
understandable: in the present calculation $q(t)$ is calculated by
solving the time-dependent Schr\"odinger equation and calculating the
expectation value from the time-dependent wave function comprising
different photon spaces. In QEDFT, the expectation
value of $q(t)$ is calculated using Eq. \eqref{max}. The dipole
moments calculated by the two approaches show some similarity. But these
quantities are also calculated in  different ways: one is solving Eq.
\eqref{r1} the other is using Eq. \eqref{hact} in the coupled orbital
and photon spaces. 

What we are really interested in is the excitation spectra
shown Fig. \ref{Na2}c. The single peak of the Na$_2$ spectrum when no
cavity is present splits into two polariton peaks in the cavity in both QEDFT and
the present calculation. The QEDFT predicts a somewhat bigger splitting
 but both approaches show a larger upper polariton peak.
The position of the disodium peak while not in a cavity seems to be at the position of
the lower polariton peak of the present calculation. This is just a coincidence:
the position of the single peak shifts to higher energies with
nonzero $\lambda$, and the present calculation also predicts upper
and lower polaritons with respect to the single peak. We will show this next.

Fig. \ref{Na2}d shows the evolution of the Rabi splitting as a
function of $\lambda$. Using the $|0\rangle$ space only, the dipole
self-interaction changes the energy of the system and changes the
position of the peak in the absorption cross section (Fig.
\ref{Na2}d) compared to the "no cavity" case (no dipole self-interaction).
By coupling the $|0\rangle$ and $|1\rangle$ photon spaces, the Rabi splitting
appears in the form of two peaks: one above and one below the single
peak of the $|0\rangle$ photon space case. Increasing $\lambda$ the
splitting is increasing as it is expected. The figure also shows the
contribution from the $|0\rangle$ component in the $|0\rangle$+$|1\rangle$ 
coupled case. For small $\lambda$ the $|0\rangle$ component is
dominant (but there is no splitting without the coupling), but for
larger $\lambda$ (see Fig. \ref{Na2}c for $\lambda=0.05$) the
$|1\rangle$ space significantly contributes; in this case to the lower
peak. Note that the width of the peaks is related to the length of the
simulation time. The present model excludes physical mechanisms of line broadening. 
More time steps at a fixed temporal resolution result in narrower, Dirac-delta
like peaks.

\begin{figure*}
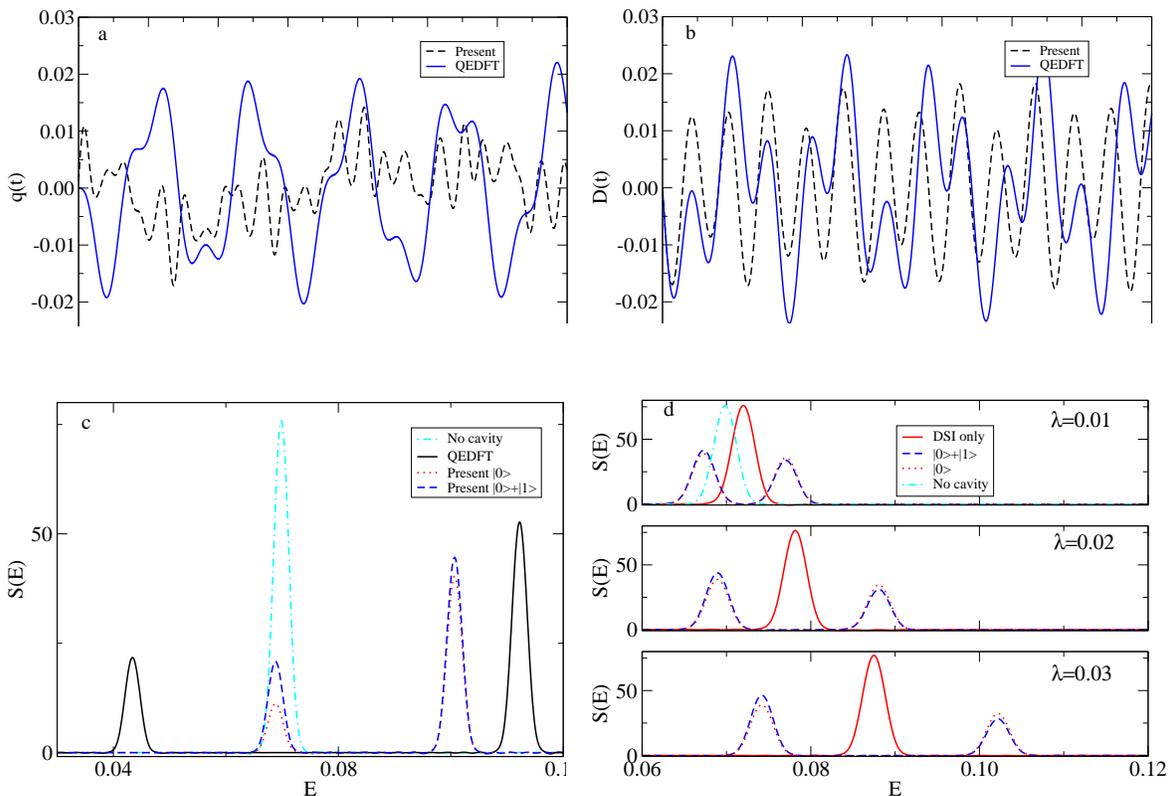

\begin{minipage}{0.95\textwidth}
\includegraphics[width=0.45\textwidth]{figure4a.eps}
\includegraphics[width=0.45\textwidth]{figure4b.eps}
\includegraphics[width=0.45\textwidth]{figure4c.eps}
\includegraphics[width=0.45\textwidth]{figure4d.eps}
\end{minipage}
\caption{QED-TDDFT-TP calculation for the Na$_2$ dimer. 
(a) Time dependence of the photon displacement coordinate.
(b) Time dependence of the dipole moment.
(c) Rabi splitting calculated by the QEDFT and by the present approach.
$|0\rangle$ is the contribution to the absorption cross section from 
the 0 photon space in a calculation using $|0\rangle+|1\rangle$ space, 
$|0\rangle+|1\rangle$ is the total absorption cross section.
$\boldsymbol{\lambda}=(\lambda,0,0)$, $\lambda=0.05$.
(d) Dependence of the Rabi splitting on $\lambda$. DSI is the dipole
self-interaction contribution. In this case the molecule is in a cavity 
but coupled only to the $|0\rangle$ photon space. The Na$_2$ dimer is
oriented  along the $x$ direction and the two atoms are d=5.8 distance apart.
$N_x=N_y=N_z=51$, h=0.5, dt=0.07, $N_t$=10000 d=5.8. $\omega=0.07$ is
used in the calculations, $N_t$=30000.}		
\label{Na2}
\end{figure*}

\subsection{Benzene and p-nitroaniline molecules}
\begin{figure}
\begin{minipage}{0.5\textwidth}
\includegraphics[width=0.3\textwidth]{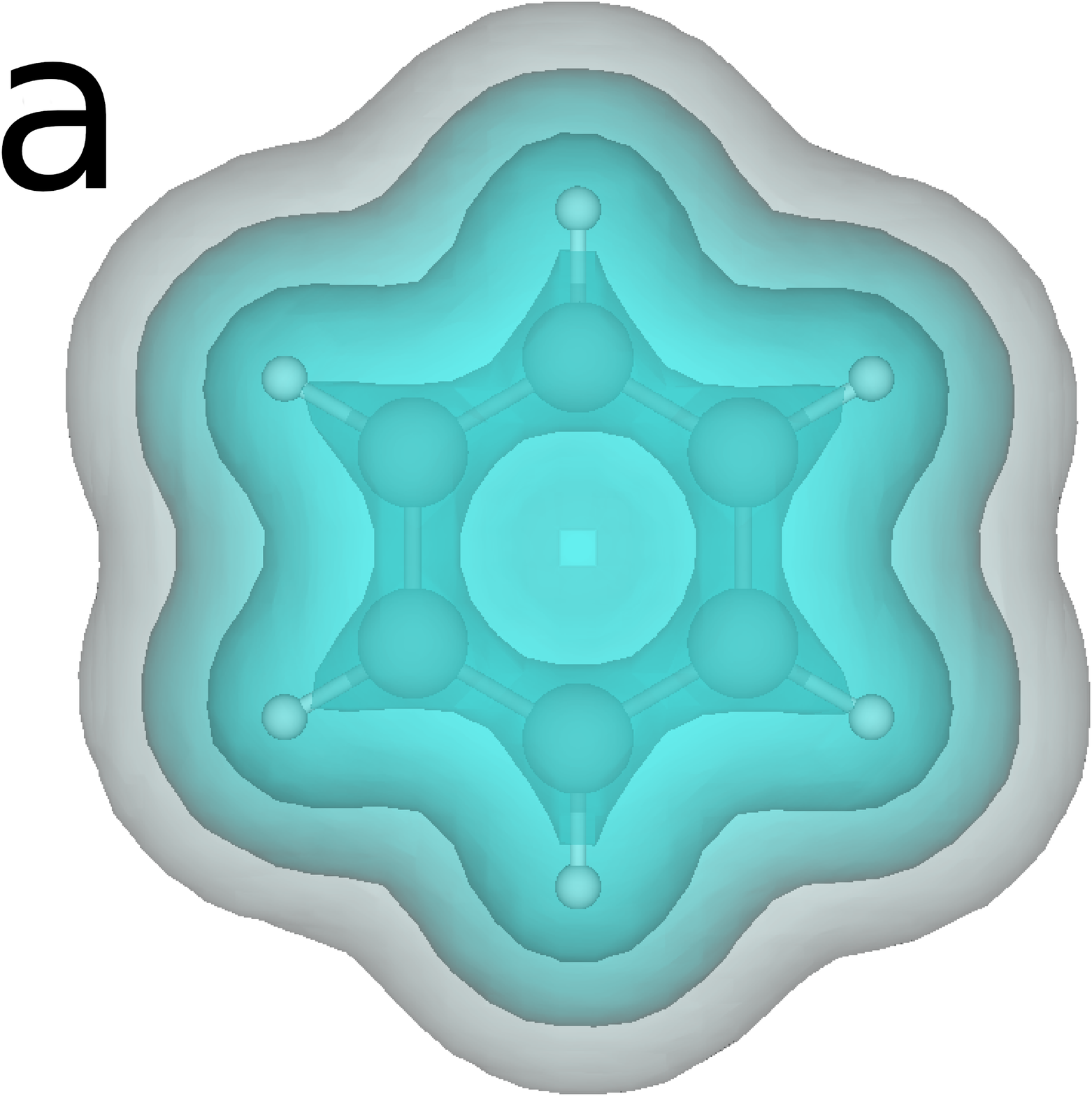}
\includegraphics[width=0.3\textwidth]{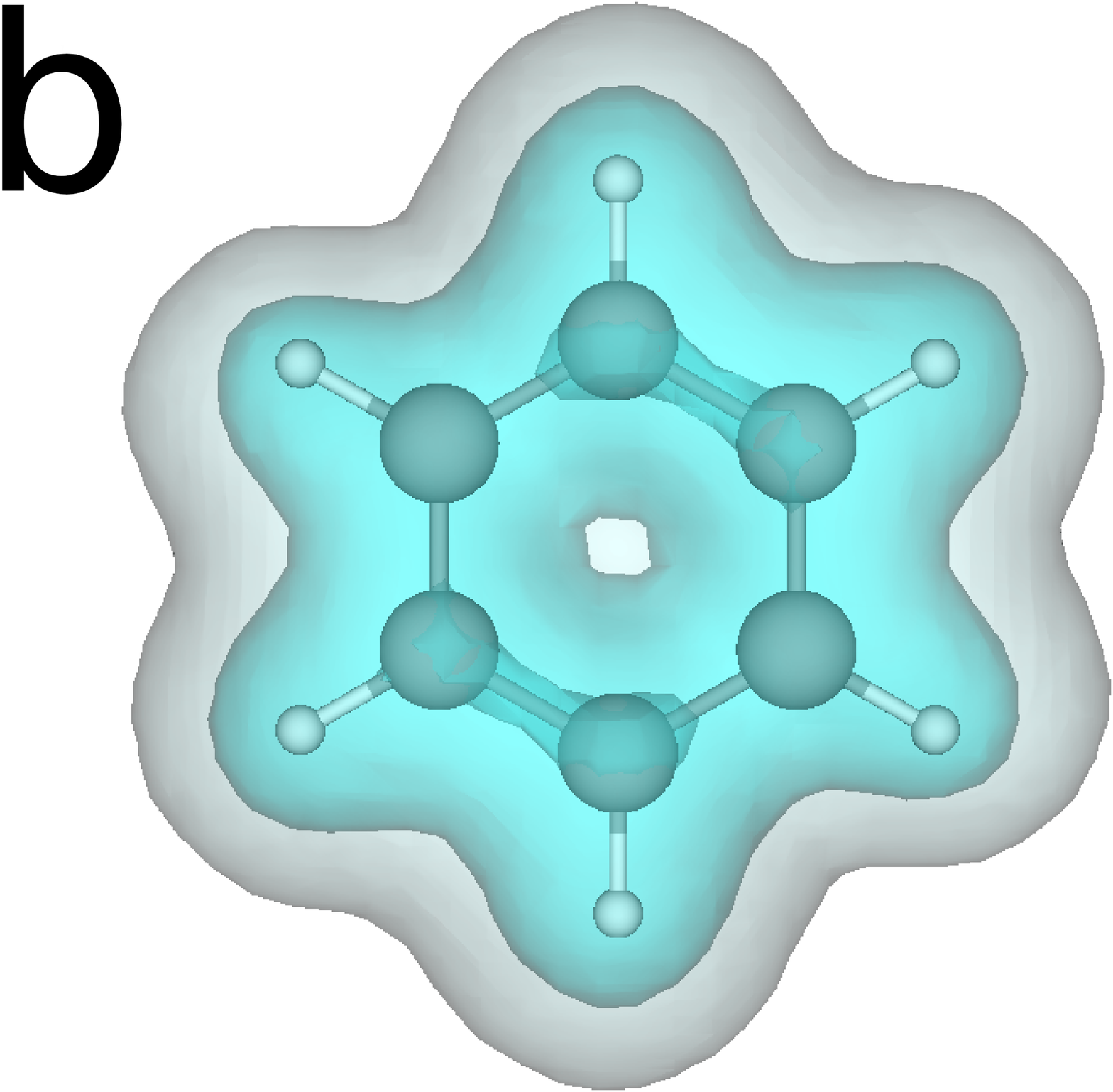}
\includegraphics[width=0.3\textwidth]{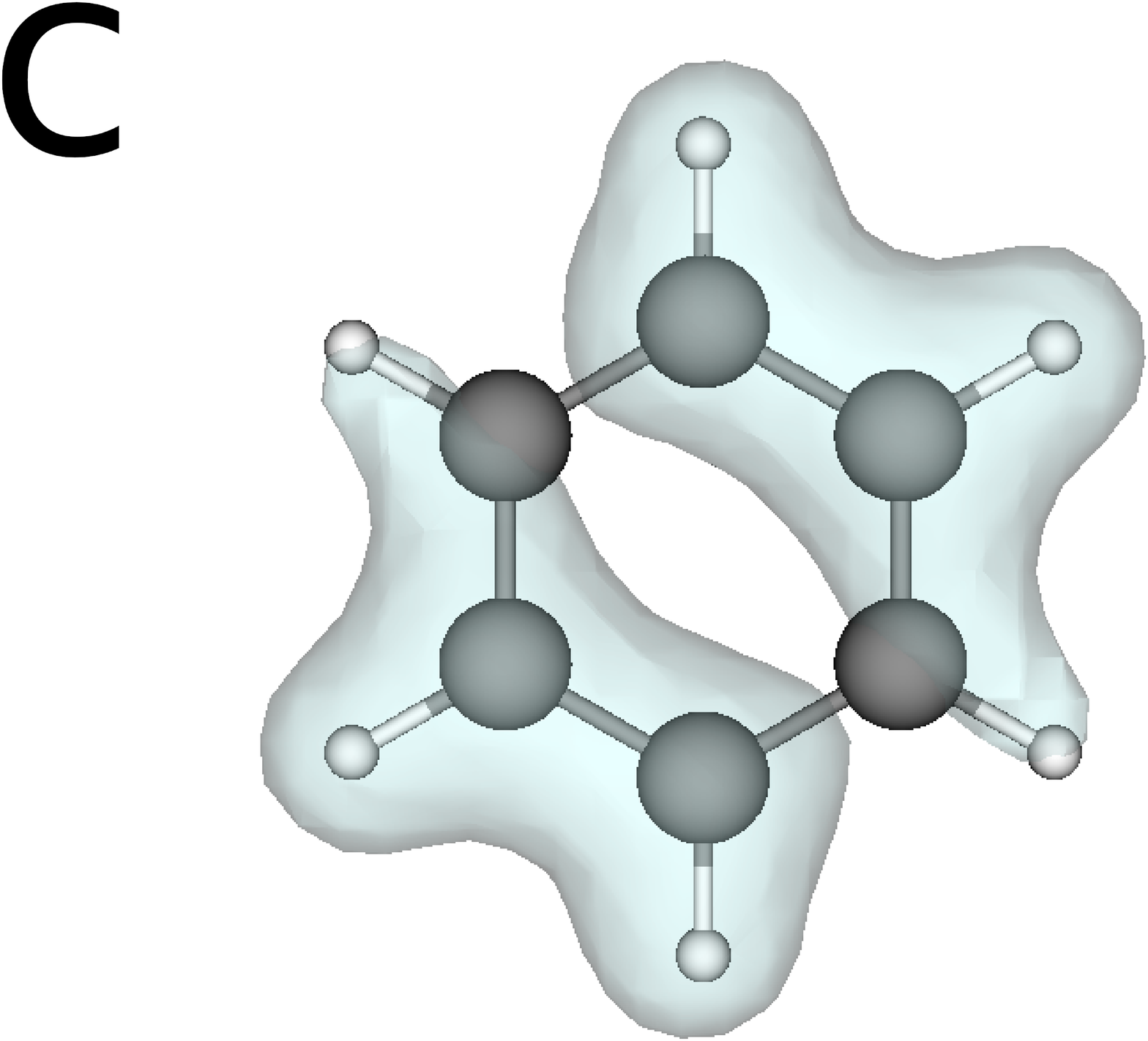}\\\quad
\includegraphics[width=0.3\textwidth]{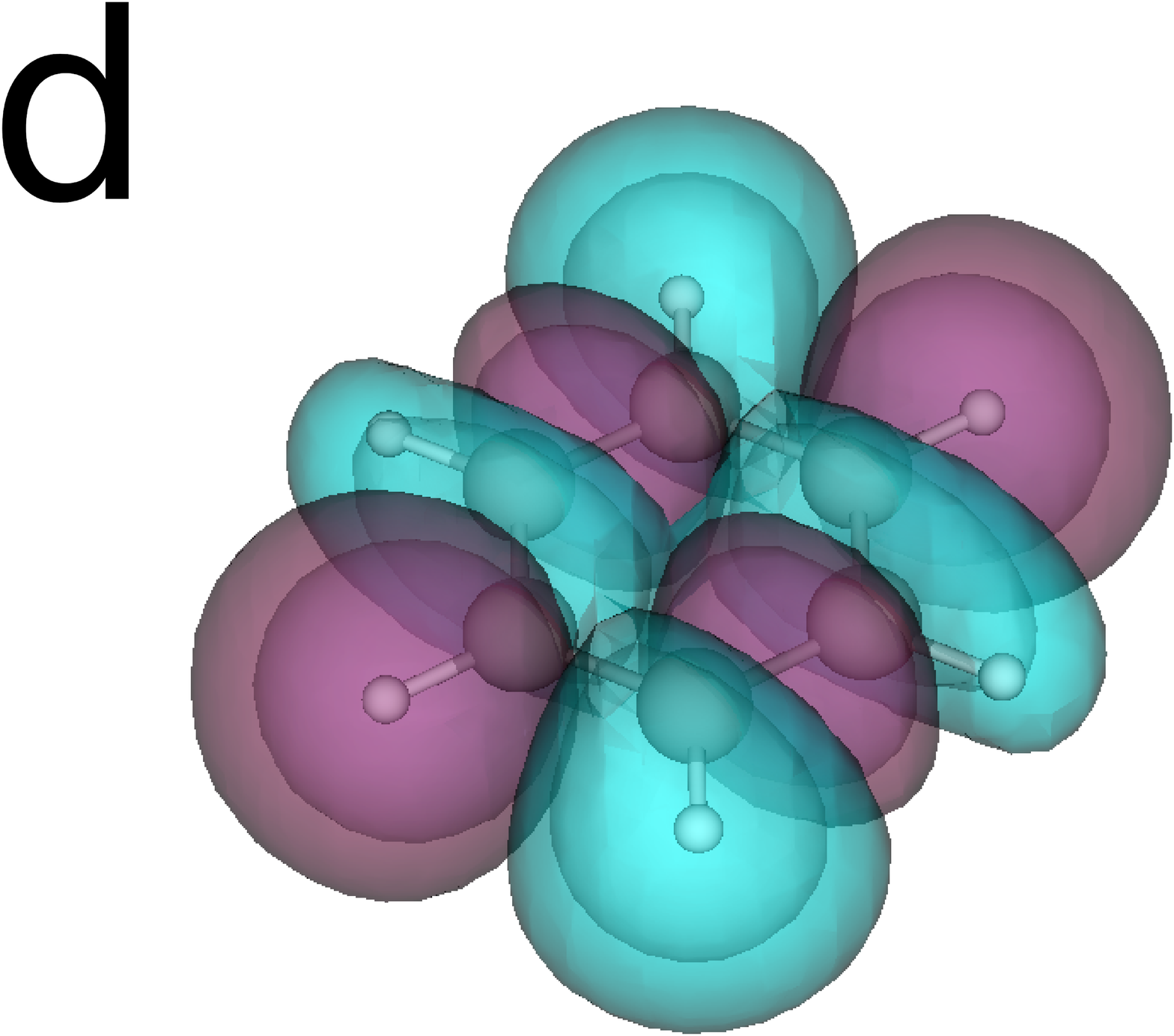}
\includegraphics[width=0.3\textwidth]{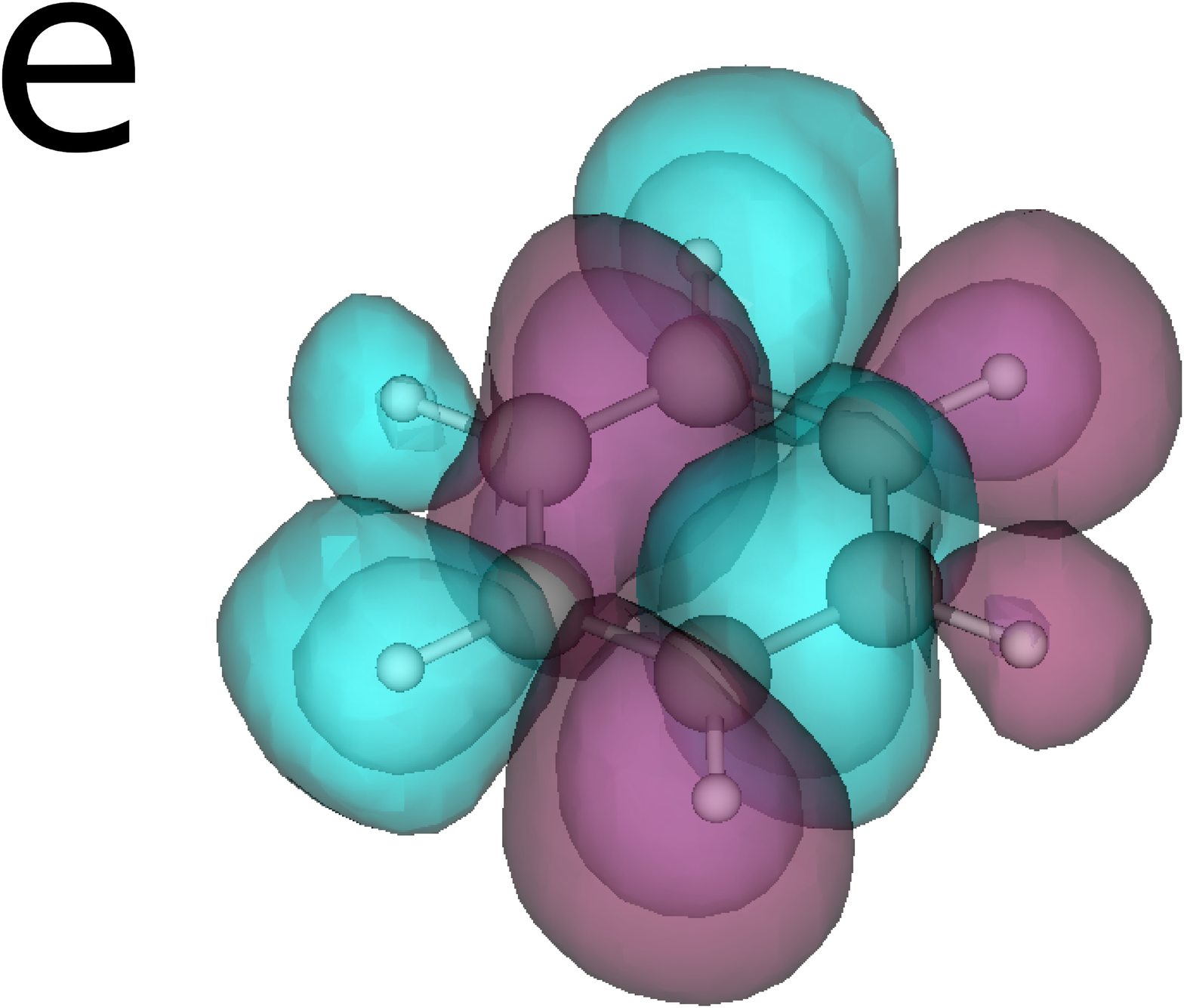}
\includegraphics[width=0.3\textwidth]{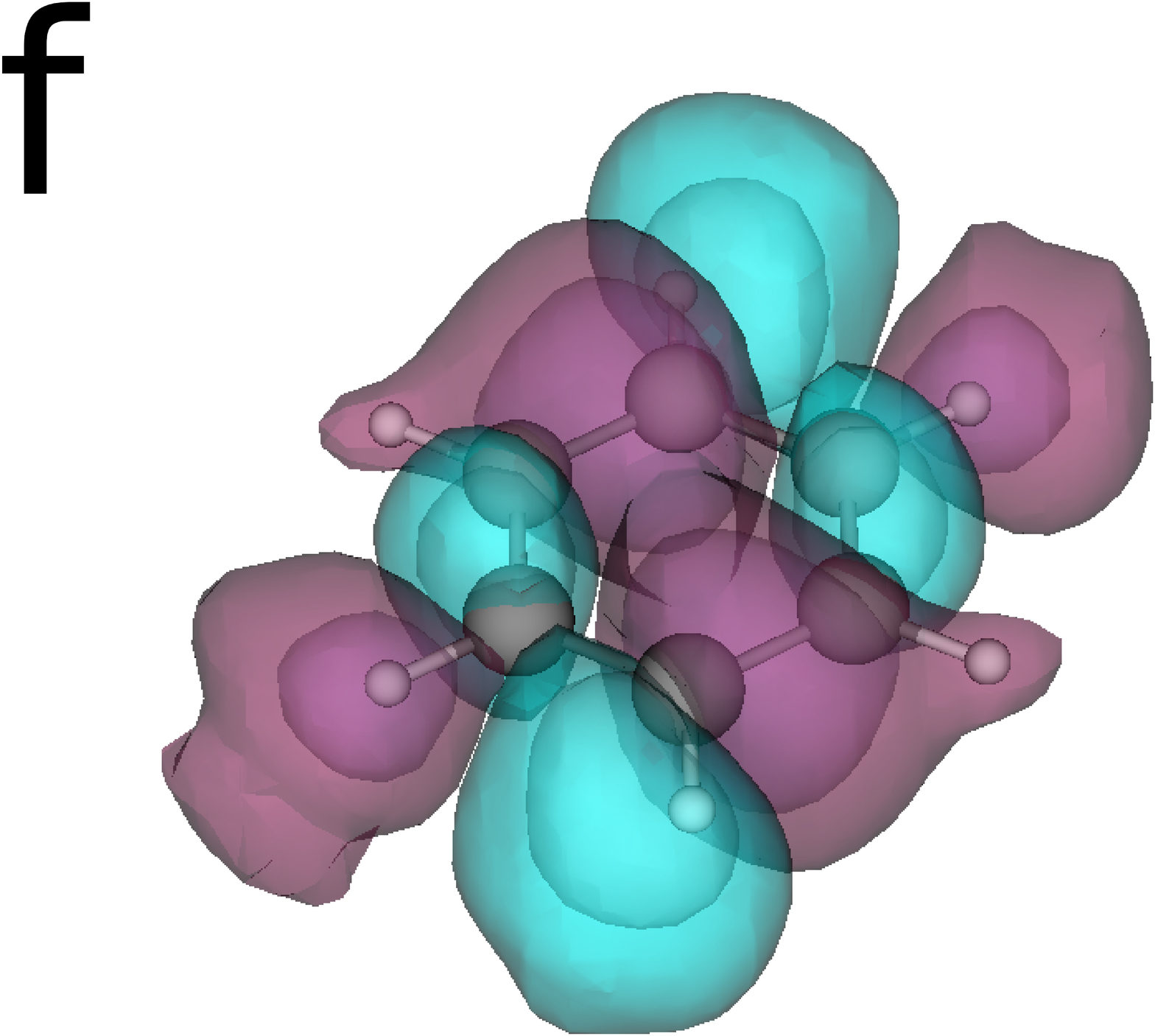}\\\quad
\includegraphics[width=0.3\textwidth]{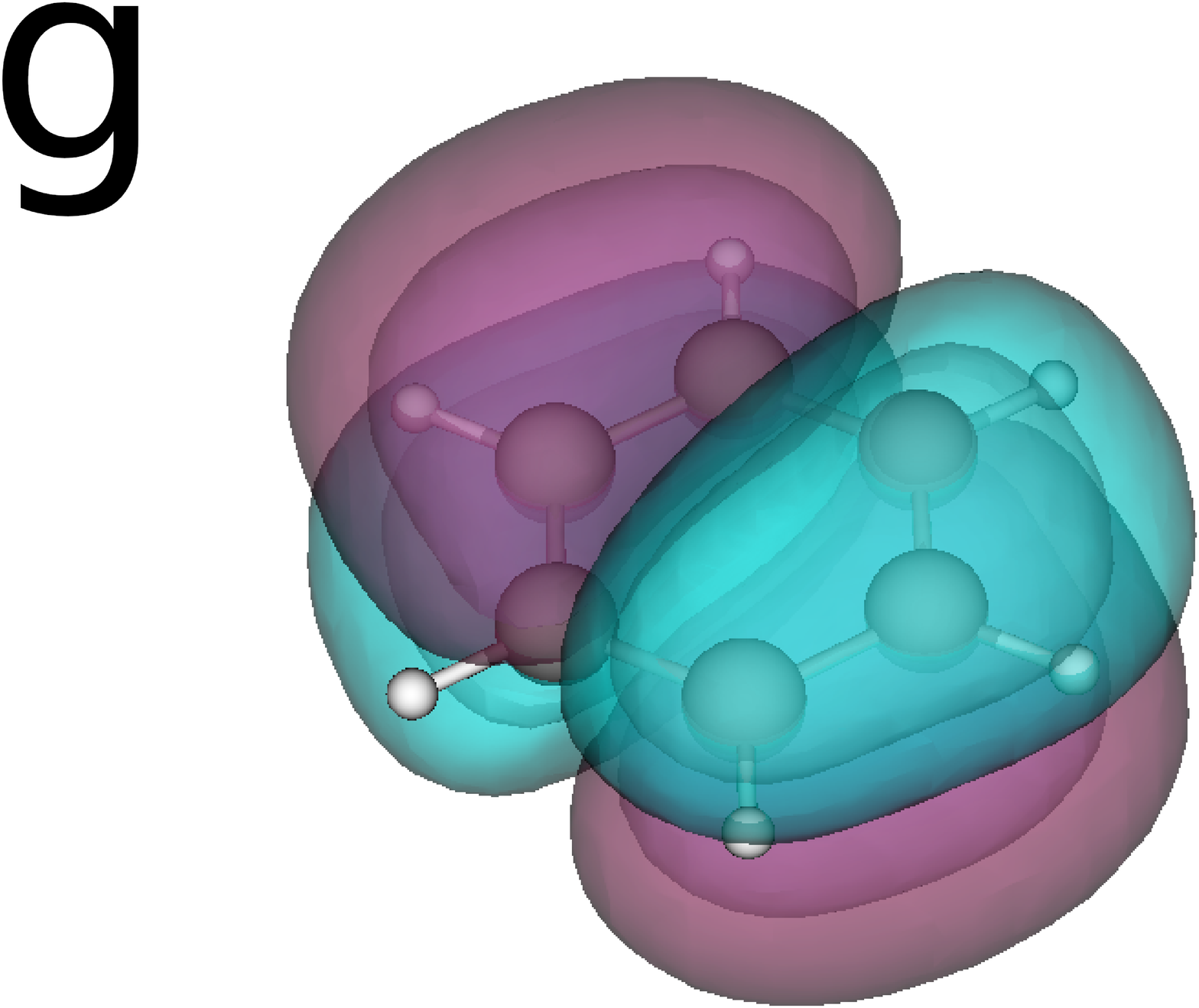}
\includegraphics[width=0.3\textwidth]{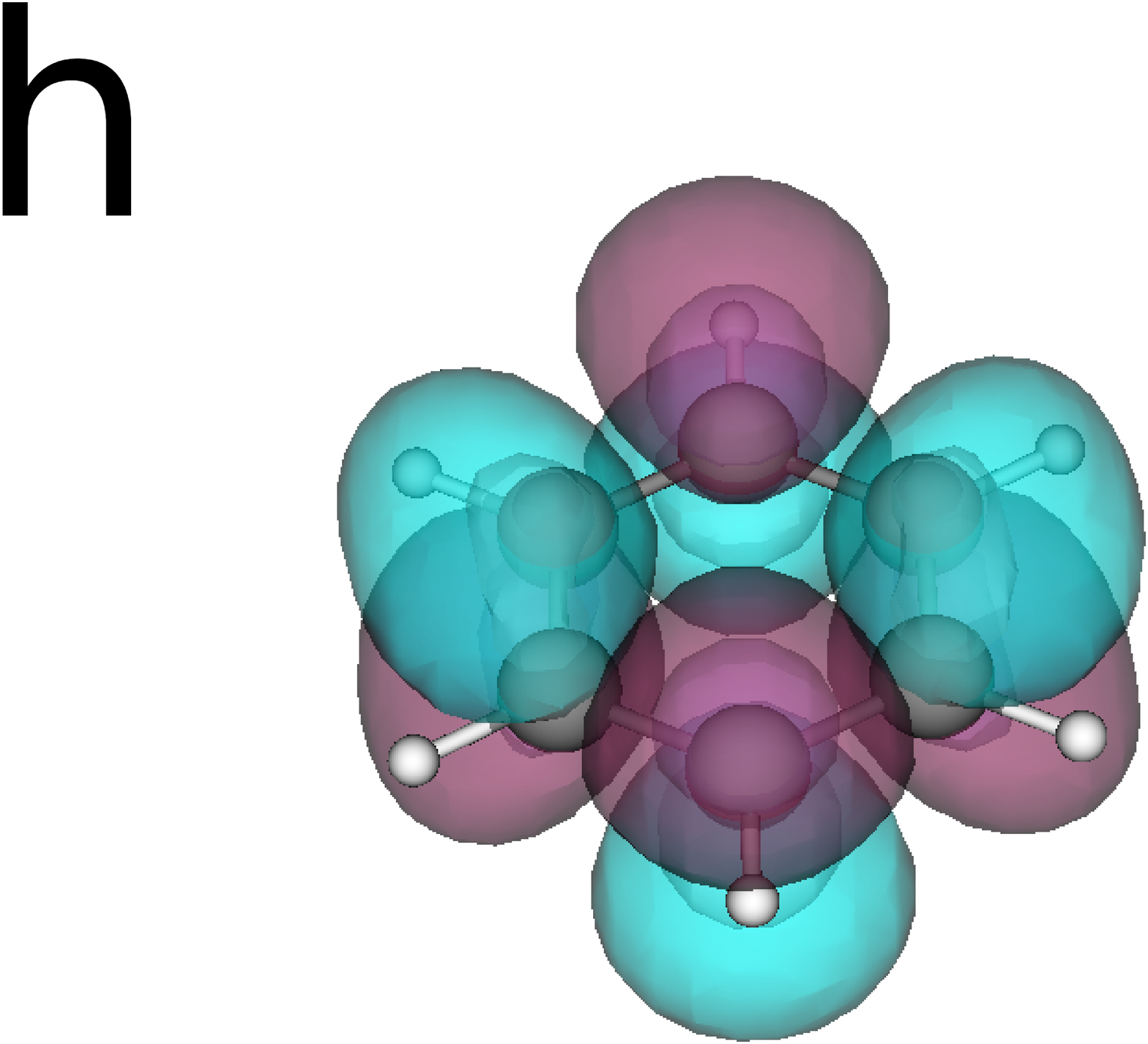}
\includegraphics[width=0.3\textwidth]{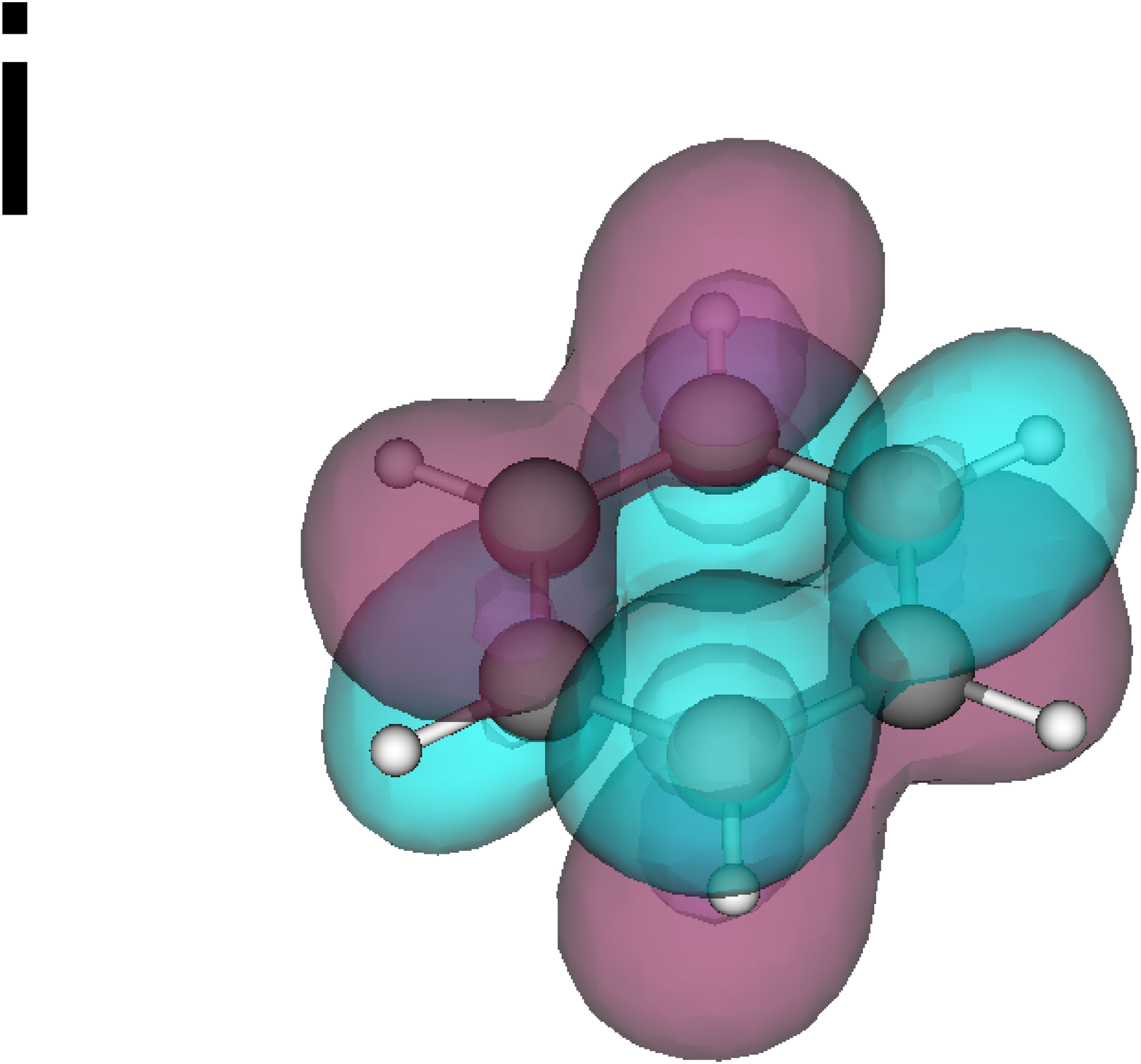}\\\quad
\end{minipage}
\caption{
Electron density ISO surfaces of benzene in the 
(a) $|0\rangle$,
(b) $|1\rangle$,
(c) $|2\rangle$
photon spaces.
The three shades of the ISO surfaces have values of 0.01, 0.001,
and 0.0001
HOMO-2 orbital isosurfaces of benzene in the 
(d) $|0\rangle$,
(e) $|1\rangle$,
(f) $|2\rangle$
photon spaces. 
HOMO orbital isosurfaces of benzene in the 
(g) $|0\rangle$,
(h) $|1\rangle$,
(i) $|2\rangle$
photon spaces. 
$N_x=N_y=N_z=51$ grid points are used with $h=0.5$ grid spacing, 
$\boldsymbol{\lambda}=(0.1,0.1,0)$, $\omega=0.5$. The darker (magenta)
color shows the positive and the lighter (cyan) color shows the
negative values.
The two shades of the isosurfaces have values of 0.0005 and 0.0002.
}
\label{benzene}
\end{figure}

In this section, we show how the electron density changes in different
photon spaces in cavities \cite{Riso2022}. The first example is a benzene molecule
(see Fig. \ref{benzene}). In this case we try a larger $\lambda$ to
couple higher photon spaces in the wave function. The $|0\rangle$,
$|1\rangle$, and $|2\rangle$ spaces are coupled and the weights of these
components are 0.94, 0.056, and 0.004. The electron density of the
benzene in the $|0\rangle$ photon space (Fig. \ref{benzene}a) is very
similar to the ground state electron density. The density in the
$|1\rangle$ photon space (Fig. \ref{benzene}b) is somewhat more compact than that in
$|0\rangle$ and a density hole appears in the middle of the ring. The
$|2\rangle$ density (Fig. \ref{benzene}c) 
is split perpendicular to the
polarization vector $\boldsymbol{\lambda}=(\lambda,\lambda,0)$. This
split is the consequence of the symmetry breaking introduced by the
$\boldsymbol{\lambda}$ dependent parts of the $H_{ep}$ term in Eq.
\eqref{hep}. 

Figs \ref{benzene}d-i show the HOMO-2 and the HOMO orbitals. The
orbitals in the $|0\rangle$ photon space remain very similar to the ground
state orbitals (the slight modification due to the dipole
self-interaction is not visible in the figures). The $|1\rangle$ and
the $|2\rangle$ components of the HOMO-2 and HOMO orbitals (Figs.
\ref{benzene}e, f, and \ref{benzene}g, i)  are significantly different
from the $|0\rangle$ component. In the case of HOMO-2 only the charge
distribution changes and the shape of the orbital components are
similar, but in the case of the HOMO  the lobes are also
different. Other orbitals (not shown) have similar shape changes.

\begin{figure}
\includegraphics[width=0.45\textwidth]{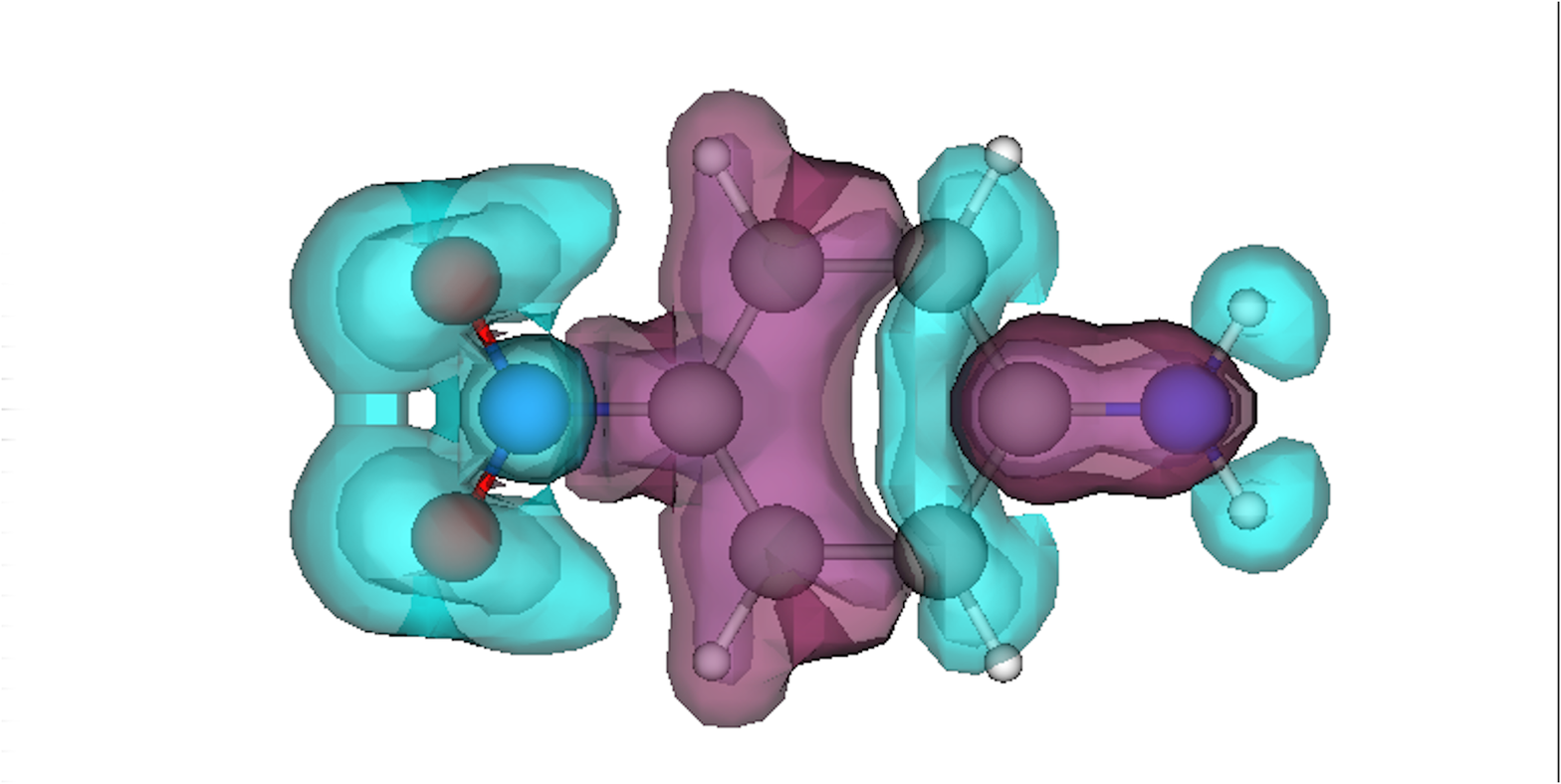}
\includegraphics[width=0.45\textwidth]{figure6b.eps}
\caption{
(top)The cavity induced density difference between
of a p-nitroaniline molecule in cavity
shown as isosurfaces.
The two shades of the isosurfaces have values of 0.0005 and 0.0002.
(bottom) Charge distribution of a p-nitroaniline molecule in cavity.
$\boldsymbol{\lambda}=(0.05,0,0)$, $\omega=0.178$ values were used in the calculation.
$N_x=N_y=N_z=55$ grid points is used with $h=0.5$ grid spacing.
The lighter (cyan) and darker (magenta) regions represent charge 
depletions and accumulations,respectively.
}
\label{PNA}
\end{figure}

The next example is a p-nitroaniline molecule which was studied by
QED-CC in Ref. \cite{PhysRevX.10.041043} and we have adapted the parameters
from that work. This is an example of a molecule with a low lying charge
transfer excitation. To measure the charge transfer one can define
\cite{PhysRevX.10.041043} 
\begin{equation}
\Delta q(x)=\int_{-\infty}^{x}dx\int_{-\infty}^{\infty}dy
\int_{-\infty}^{\infty}dz 
\Delta\rho(x,y,z),
\end{equation}
\begin{equation}
\Delta \rho(x,y,z)=\rho^{cavity}(x,y,z)-\rho^{no cavity}(x,y,z).
\end{equation}
$\Delta q(x)$ measures the amount of charge moved in the $x$ direction -- the
principle axis of the molecule. The density difference induced
in the molecule by the cavity is shown in Fig. \ref{PNA}. The charge
redistribution is very similar to what is shown in Ref. \cite{PhysRevX.10.041043}.
Electrons transfer from the nitro (NO$_2$) side to the aniline (NH$_2$) side.  The
$\Delta q(x)$ function shown in Fig. \ref{PNA} is also very similar to QED-CC
calculation in Ref. \cite{PhysRevX.10.041043}. 

\begin{figure}
\includegraphics[width=0.45\textwidth]{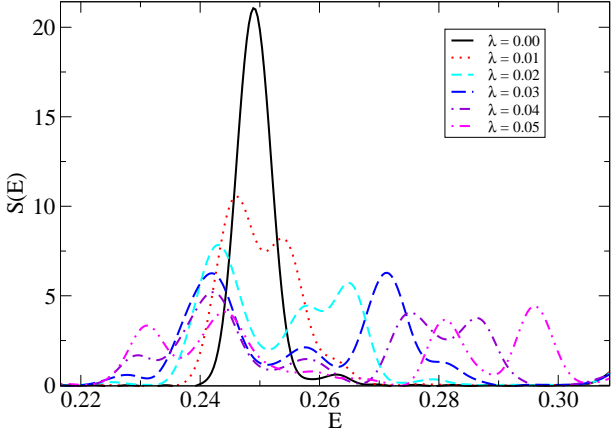}
\caption{Rabi splitting for benzene for different $\lambda$ values.
$N_x=N_y=N_z=55$ grid points is used with $h=0.5$ grid spacing. The
time step is 0.05 and the total propagation time is 1000. 
$\boldsymbol{\lambda}=(0,\lambda,0)$ and $\omega=0.25$ was used in the
calculations.}		
\label{brabi}
\end{figure}

We have also calculated the optical absorption spectrum of the benzene
molecule in a cavity by time propagating the dipole moment after an
initial delta kick perturbation.
The Rabi splittings for benzene is shown in Fig. \ref{brabi}. 
The calculated splittings are 0.22 (0.33), 0.63 (0.69), 0.81 (1.02), 1.03
(1.35) for $\lambda=0.01, 0.02, 0.03, 0.04$, respectively. The numbers in
the parenthesis are from Ref.  \cite{doi:10.1021/acsphotonics.9b00768}.
Just as in the case of the Na$_2$ cluster our QED-TDDFT-TP approach gives 
somewhat smaller Rabi splitting than the QEDFT approach, but the overall
trend, a direct proportionality between the splitting and $\lambda$, is the same.

\begin{figure*}
\begin{minipage}{0.95\textwidth}
\includegraphics[width=0.45\textwidth]{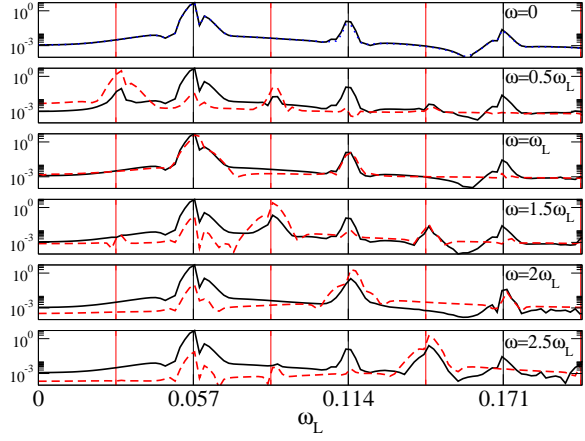}
\end{minipage}
\caption{High harmonics spectrum of the HF molecule (black lines). The
red dashed lines show the contribution from the $\vert 1\rangle$
photon space. The blue dotted line shows the high harmonics of the HF
molecule without  the cavity. 
The solid vertical lines show the position of $n\omega_L$, the dashed
vertical lines are at $(n+1/2)\omega_L$ (where $n$ is an integer). 
The axis of the molecule molecule is in the $x$ direction, the
H-F distance is 1.7229, $N_x=N_y=N_z=51$, $h=0.5$,
$\boldsymbol{\lambda}=(0.05,0,0)$, $\Delta t=0.05$ and $\omega=0.081$ is used. 
The laser is polarized in the $x$ direction and the laser parameters are
$\omega_L$=0.057 (800 nm wavelength), $T_L=2/\omega_L$ and $E_x$=0.005.}		
\label{hhg}
\end{figure*}

\subsection{High harmonic generation}
As a final example we present the high harmonic spectrum of the HF
molecule with and without a cavity. The molecule is excited with a continuous laser pulse,
\begin{equation}
E(\mathbf{r},t)=(E_x,0,0)\sin(\pi t/(6 T_L))^2
\sin(\omega_L t)
\end{equation}
and after 30 laser cycles the high harmonic spectrum is calculated using the dipole acceleration:
\begin{equation}
I(\omega)=\left|\int_0^T{\partial^2 \mathbf{D}(t)\over \partial t^2}{\rm e}^{-i\omega t} dt\right|^2.
\end{equation}
Fig. \ref{hhg} shows the calculated high harmonics. We are mostly
interested in the first few harmonics and use a relatively small computational box 
to speed up the calculations. To resolve more  harmonics one
needs a larger box, which increases the computational cost.

If no cavity is present, the harmonic spectrum of the HF molecule oriented
in the direction of the polarization of the laser consists of peaks at
$n\omega_L$ (see Fig \ref{hhg} for the $\omega=0$ case). Since HF does not have inversion symmetry, 
both the even and odd harmonics are present. Placing the molecule in a
cavity using the $\vert 0\rangle$ photon space only, the high harmonic
spectrum of the HF molecule does not change (the solid black and
dotted blue lines are barely distinguishable in Fig \ref{hhg}). 

Using the $\vert 0\rangle+\vert 1\rangle$ photon spaces new peaks appear 
in the harmonic spectrum at $\omega+n\omega_L$ positions. For example
if $\omega=0.5 \omega_L$  then the new peaks are at $0.5\omega_L+
n\omega_L$ (Fig. \ref{hhg}). These peaks have significant intensity, and the
peaks are present at weaker and stronger couplings as well. If
$\omega=n\omega_L$ then the new (cavity induced) peaks coincide with
peaks of the laser and the spectrum  remains very similar to the harmonic
spectrum without a cavity. Fig. \ref{hhg} also shows
the contribution from the $|1\rangle$ photons spaces. The dipole
moment is the sum of the components from the $\vert 0\rangle$ and 
$\vert 1 \rangle$ photon spaces, and the harmonic spectrum is a result of an
interference of these terms. Fig \ref{hhg} shows that in the case of
$\omega=(k+1/2)\omega_L$ (k=0,1,2{\ldots}) the $\vert 1\rangle$ component 
has pronounced peaks at $\omega$,$\omega+\omega_L$ and
$\omega+2\omega_L$. The $\vert 1\rangle $ photon space not only
induces new peaks, but significantly contributes to their intensity as
well.

\section{Summary}
We have developed and implemented a QED-TDDFT-TP approach that uses a direct product 
of a Fock space for light and a real-space grid for electrons. We have
tested the approach by comparing it to accurate QED-SVM and QED-CC
calculations and the results are very promising. The coupling strength
and light frequency dependence of energies calculated by the present
approach is very similar to the QED-SVM and QED-CC results. 

One of the main advantages of the approach is that it preserves the quantized
photon spaces giving direct access to non-classical photon observables.
A further important advantage is that as the electron 
and the photon coordinates are separated one can asses the
contribution to physical quantities  from different photon spaces.

The light-matter coupling strongly modifies the molecular orbitals, and
the orbitals have very different shapes in different photon spaces.
This also leads to charge transfer and redistribution, and our results
are similar to the QED-CC  calculation \cite{PhysRevX.10.041043}.

We have also studied the Rabi splitting in molecules in optical
cavities and the results show that the splitting is caused by
the presence of light states: there is no splitting  without the
presence of $|0\rangle$ and $|1\rangle$ photon spaces. 

The approach can also be used to study high harmonic generation in
strongly coupled light-matter coupled systems. There are several recent experimental
and theoretical works studying the influence of polaritons on
nonlinear optical processes \cite{Kockum2017,doi:10.1021/acs.nanolett.6b02567,doi:10.1021/acsphotonics.7b00305,
https://doi.org/10.1002/adom.201801682,https://doi.org/10.48550/arxiv.2203.00691}.
We plan to extend the present study to describe the high
harmonic generation for more complex molecules. We also plan to calculate the nonlinear susceptibilities 
of molecules in cavities using combining the present approach with the
method described in Ref. \cite{doi:10.1063/1.4749793}. 

The tensor product of the real-space and Fock-space increases by the
number of Fock spaces. For experimentally realizable strong coupling
one needs $N_F$=2 or $N_F$=3 Fock space dimensions for a single photon frequency. This
increases the cost of the calculation in time and memory requirements
by a factor of $N_F$. If there are more than one photon frequency is
important then
the calculation cost increases by $N_F^{N_p}$ where $N_p$ is the number of
relevant photon modes. For larger $N_p$ this  will be prohibitively expensive and
one needs alternatives, such as the QEDFT approach, which can handle any
number of photons.

The increased dimensionality is not the only complicating factor. The
Hamiltonian \eqref{hact} (except from a trivial $n\omega$ shift) acts
the same way on every real-space component $\phi_{mn}(\mathbf{r},t)$ for a
given photon number. If there would be no coupling in the Hamiltonian
then one would have the same set of  $\phi_{mn}(\mathbf{r},t)$,
($m=1,{\ldots}, N_{occ}$) orbitals  in each $|n\rangle$ photon space.
In the self-consistent iterative diagonalization of the nonlinear 
Hamiltonian, one has to carefully mix the density so that the photon 
components are gradually optimized. The process is very sensitive to
the coupling (the off-diagonal part of the Hamiltonian), because the 
coupling makes the orbitals different in different photon spaces.
Sudden changes in the density leads to slow  damped oscillatory convergence.
Future work is necessary on optimizing the self-consistent solution
for this tensor product
approach to have efficient calculations.

The development of accurate exchange-correlation functionals are also
important for future calculations. In the present work we have
compared our calculations to QED-SVM and QED-CC results, and these
approaches may help in developing appropriate functionals for QED-TDDFT-TP.

\begin{acknowledgments} 
This work has been supported by the National Science
Foundation (NSF) under Grant No. IRES 1826917.
J.M., A.A. and D.P. equally contributed to this work.
\end{acknowledgments}

\section*{Data Availability Statement}
The data that support the findings of this study are available
from the corresponding author upon reasonable request.

\section*{AUTHOR DECLARATIONS}
\par\noindent
{\bf Conflict of Interest}

The authors have no conflict of interest to disclose.

\appendix

\section{Photon space}
\label{boson}
In this Appendix we give a brief overview of the properties of the Fock
space and occupation number basis.
The photon Hamiltonian (from Eq. \eqref{hep})
\begin{equation}
H_{p}=\sum_{\alpha=1}^{N_p}{1\over 2}\left(p_{\alpha}^2+\omega_{\alpha}^2 q_{\alpha}^2\right)
=
\sum_{\alpha=1}^{N_p} \omega_{\alpha}\left(
\hat{a}_{\alpha}^{+}\hat{a}_{\alpha}
+\frac{1}{2}\right),
\end{equation}
where 
\begin{equation}
q_\alpha={\frac{1}{
\sqrt{2\omega_\alpha}}}(\hat{a}_\alpha+\hat{a}^+_\alpha) \ \ \ \ 
p_\alpha={\frac{1}{i
\sqrt{2\omega_\alpha}}}(\hat{a}_\alpha-\hat{a}^+_\alpha).
\end{equation}
Denoting the vacuum state as $|0\rangle$,  any
eigenstate of $\hat{a}_{\alpha}^{+}\hat{a}_{\alpha}$ can be calculated
by multiple applications of the creation operators 
\begin{equation}
|n_{\alpha}\rangle=
{1\over\sqrt{n!}} (\hat{a}_{\alpha}^{+})^n |0\rangle.
\end{equation}
This is called the occupation number representation, $n_{\alpha}$ defines the 
excited state of the $\omega_{\alpha}$ photon mode.
The bosonic operators satisfy the commutation relations
\begin{equation}
[\hat{a}_{\alpha},\hat{a}_{\alpha'}]=
[\hat{a}_{\alpha}^+,\hat{a}_{\alpha'}^+]=0, \ \ \ \ \
[\hat{a}_{\alpha},\hat{a}_{\alpha'}^+]=\delta_{\alpha,\alpha'}\hat{1}
\end{equation}
where $[a,b]=ab-ba$.

Now one can define an occupation number basis
\begin{equation}
\chi_{\vec{n}}=\vert n_1,n_2,{\ldots},N_p\rangle=
{1\over\sqrt{n_1!{\ldots} n_{N_p}!}} 
(\hat{a}_{1}^{+})^{n_1}{\ldots} 
(\hat{a}_{N_p}^{+})^{n_{N_p}}|0\rangle
\label{oc}
\end{equation}
where $n=n_1+{\ldots} N_{p}$. 
In this abstract space representation the symmetry requirement for
bosons is satisfied by allowing that mode $i$ is occupied by $n_i$
photons. For the product state in Eq. \eqref{oc} the restriction to the symmetric 
subspace is tacitly assumed, that is one has to symmetrize the states
for identical harmonic oscillators.

\section{QEDFT}
\label{Rubio}

In the approach of Refs.
\cite{Schafer4883,Ruggenthaler2018,Flick3026,
doi:10.1021/acs.jpclett.0c03436,doi:10.1021/acsphotonics.9b00768}
two coupled equation are solved
\begin{equation}
i \hbar \frac{\partial}{\partial t} \varphi_{i}(\mathbf{r},
t)=(-{1\over 2}\nabla^2+V_{KS}(\mathbf{r},t)+V_P(\mathbf{r},q,t)) 
\varphi_{i}(\mathbf{r}, t)
\label{r1}
\end{equation}
\begin{equation}
\left(\frac{\partial^{2}}{\partial t^{2}}+\omega_{\alpha}^{2}\right)
q_{\alpha}(t)=-\frac{j_{\alpha}(t)}{\omega_{\alpha}}+\omega_{\alpha}
\boldsymbol{\lambda}_{\alpha} \cdot \mathbf{R}(t)
\label{max}
\end{equation}
where 
$\mathbf{R}(t)=\int d \mathbf{r} e \mathbf{r} \rho(\mathbf{r}, t)$.
The photon exchange potential is 
\begin{equation}
V_{\mathrm{P}}(\mathbf{r},q,t)
=  \sum_{\alpha=1}^{M}\left(\int d
\mathbf{r}^{\prime} \boldsymbol{\lambda}_{\alpha} \cdot
\mathbf{r}^{\prime} \rho\left(\mathbf{r}^{\prime},
t\right)-\omega_{\alpha} q_{\alpha}(t)\right)
\boldsymbol{\lambda}_{\alpha} \mathbf{r}
\end{equation}

The photon is propagated as 
\begin{equation}
\left({\partial^2\over \partial
t^2}+\omega_{\alpha}^2\right)q_{\alpha}(t)=\omega_{\alpha}\boldsymbol{\lambda}_{\alpha}
\mathbf{R}(t)
\end{equation}

\section{Stochastic variational approach}
In this case the full Coulombic Hamiltonian is used, the Hamiltonian of an $N$ electron system interacting 
with a Coulomb interaction and confined in 
an external potential $V_c$ is
\begin{equation}
    H_e=-\sum_{i=1}^N {{\nabla^2_i}\over 2 m_i}+\sum_{i<j}^N
    {q_iq_j\over \vert \mathbf{r}_i-\mathbf{r}_j\vert}+\sum_{i=1}^N V_c(\mathbf{r}_i),
\end{equation}
where $\mathbf{r}_i,q_i$, and $m_i$ are the coordinate, charge, 
and mass of the $i$th particle ($m_i=1$ for electrons in atomic units).

Introducing the shorthand notations  $\vec{r}=(\mathbf{r}_1,...,\mathbf{r}_N)$, 
the variational trial wave function is written as a linear
combination of  products of spatial and photon
space basis functions
\begin{equation}
\Psi(\vec{r})=\sum_{n}\sum_{k=1}^{K_{n}}
    c_k^{n}\psi_k^{n}(\vec{r})|n\rangle.
    \label{pwfM}
\end{equation}

The spatial part of the wave function is expanded into ECGs 
for each photon state $|n\rangle$ as
\begin{equation}
    \psi_k^{n}(\vec{r})={\cal A}\lbrace {\rm e}^{-{1\over 2}\sum_{i<j}^N
    \alpha_{ij}^k(\mathbf{r}_i-\mathbf{r}_j)^2-{1\over 2}\sum_{i=1}^N \beta_i^k(\mathbf{\mathbf{r}}_i-\mathbf{s}_i^k)^2}
    \chi_S\rbrace
\label{bf}
\end{equation}
where ${\cal A}$ is an antisymmetrizer, $\chi_S$ is the $N$ electron spin function (coupling the spin to $S$), and 
$\alpha_{ij}^k$,$\beta_i^k$ and $\mathbf{s}_i^k$ are nonlinear parameters.
The basis functions are optimized as described in Ref. \cite{PhysRevLett.127.273601}.

\section{Absorption cross section}
\label{abs}
The polarizability tensor \( \boldsymbol{\alpha}(\boldsymbol{\omega})
\) can be calculated  by time propagation of the electron orbitals states.
The initial state is first perturbed by a delta kick potential
\begin{equation}
V_{ext}=-e \mathbf{r} \cdot \mathbf{k}_{i} \delta\left(t_{0}\right),
\end{equation}
where \( \mathbf{k}_i \) is the 
electric field in the $i$ direction ($i=x,y,z$. The magnitude of the
electric field  should be sufficiently small to initiate a linear-response.

The polarizability tensor in frequency space is defined as the Fourier Transform 
of the time-dependent dipole moment 
\begin{equation}
 \alpha_{ij}(\omega)=\frac{1}{k_i} \int_{0}^{\infty}
 d t\left[D_j(t)-D_j\left(t_{0}\right)\right] e^{-i \omega t} .
\end{equation}
The time-dependent dipole moments can be calculated from the time-dependent electronic
density  \( (\rho(\boldsymbol{r}, t))) \) :
\begin{equation}
 \mathbf{D}(t)=\int \mathbf{r} \cdot \rho(\boldsymbol{r}, t) d \mathbf{r}
\end{equation}
The imaginary part of the dynamic
polarizability tensor can be related to the oscillator strength of
each transition \cite{ullrich}
\begin{eqnarray}
 \frac{1}{3} \operatorname{Tr}(\Im m
 \boldsymbol{\alpha}(\omega))&=&\frac{\pi}{3} \sum_{n=1}^{\infty}
 \sum_{\nu=1}^{3}\left|\left\langle\Psi_{0}\left|\hat{r}_{\nu}\right|
 \Psi_{n}\right\rangle\right|^{2}
 \delta\left(\omega-\Omega_{n}\right)\nonumber \\
 &=&\sum_{n=1}^{\infty} \frac{\pi}{2
 \Omega_{n}} f_{n} \delta\left(\omega-\Omega_{n}\right),
\end{eqnarray}
where the \( \left|\Psi_{n}\right\rangle \) is the wave function of
the \( n \)-th excited state, \( \Omega_{n} \) is the corresponding
excitation energy, and \( f_{n} \) is the oscillator strength belonging
to the transition probability for the excitation. The dynamic polarization tensor \(
 (\boldsymbol{\alpha}(\omega)) \) and the transition dipole moment are
 now connected as \(
 \left(\mathbf{D}_{n 0}=\left\langle\Psi_{0}|\hat{\boldsymbol{r}}|
 \Psi_{n}\right\rangle\right) . \)
 
From the imaginary part of the dynamic polarizability
tensor one can extract the photo-absorption cross-section: 
\begin{equation}
 \sigma(\omega)=\frac{4 \pi \omega}{3 c} \operatorname{Tr}\left[\Im
 m(\boldsymbol{\alpha}(\omega)\right]=\frac{2 \pi^{2}}{c} S(\omega)
\end{equation}
where \( c \) is the speed of light and \(
S(\omega)=\sum_{n=1}^{\infty} f_{n}
\delta\left(\omega-\Omega_{n}\right) \) is the dipole spectral function.

A damping function is  added to the Fourier transform of 
the photo-absorption spectrum for finite propagation times.
The damping function introduces an artificial decay of
the excited population to  smooth spectrum. We use a
Gaussian damping function \( \left(e^{-\eta^{2} t^{2}}\right) \) and
the parameter \( \eta \) is chosen to make the damping very small 
at the end of propagation.


\end{document}